\documentclass[12pt]{iopart}
\usepackage{graphicx}
\usepackage{iopams}
\usepackage{psfrag} 
\usepackage{fancyhdr}
\begin{document}
\date{\today}
\title{Electrostatic correlations: from Plasma to Biology}
\author{\bf Yan Levin} 
\address{\it Instituto de F\'{\i}sica, Universidade Federal
do Rio Grande do Sul\\ Caixa Postal 15051, CEP 91501-970, 
Porto Alegre, RS, Brazil\\ 
{\small levin@if.ufrgs.br}}

\begin{abstract}
 
Electrostatic correlations play an important role in physics, chemistry
and biology.
In plasmas they result in thermodynamic instability similar to the liquid-gas
phase transition of simple molecular fluids.  For charged colloidal
suspensions the electrostatic correlations are responsible for screening
and colloidal charge renormalization. In aqueous solutions
containing multivalent counterions they
can lead to charge inversion and flocculation.  
In biological systems the  correlations
account for the organization of cytoskeleton and the 
compaction of genetic material. 
In spite of their ubiquity, the true importance of electrostatic
correlations has become fully appreciated only quite recently.  
In this paper, I will review the thermodynamic consequences of
electrostatic correlations in a variety of systems ranging from
classical plasmas to molecular biology.

\end{abstract}
\maketitle
\bigskip
\pagestyle{fancy}
\lhead{\thepage}
\rhead{Electrostatic correlations: from Plasma to Biology}
\cfoot{}
\tableofcontents
\newpage

\section{Introduction}

Although the liquid state
theorists have become quite accustomed to look at the correlation
functions for simple and complex fluids, many of the thermodynamic 
consequences 
of correlations have not been fully appreciated.  To some extent this
is the result of the success of mean-field theories for simple
fluids.  The strategy of applying the  mean-field approximation to the
many body problems goes all the way to the pioneering work of van der
Waals on liquid-gas phase separation~\cite{Wa73}.  The success
of this theory and its physical transparency has set the stage
for future applications of mean-field ideas.  These came 
in the form of the Curie-Weiss theory of magnetism~\cite{We07} 
and the Gouy-Chapman~\cite{Go10,Ch13} theory of diffuse ionic layers.  

There are, however, some very familiar  systems in which the 
mean-field contribution to the free energy is identically  zero.  One
such system is the classical two component plasma of positive
and negative ions. In order for the thermodynamic limit to exist,
charge neutrality constraint must be imposed.  However,
for a bulk charge-neutral system the average electrostatic potential
is zero, which means that the mean-field contribution to the
total free energy vanishes.  Thus, the electrostatic 
free energy
of a two component plasma is entirely due to 
positional correlations between the
positive and negative ions. At low temperatures these 
correlations become so 
strong as to lead to  phase transition in which the plasma
separates into two coexisting high and low density phases~\cite{FiLe93,LeFi96}.

Polar fluids provide another example of a system in which the
electrostatic correlations strongly affect the thermodynamics.
Perhaps the simplest model of a polar fluid
is a system of dipolar hard spheres $(DHS)$. The phase structure
of $DHS$ is quite interesting and deserves a 
separate review~\cite{GrDi94,GrDi96,GrDi97}.
Here we shall confine our attention to the low density disordered
fluid phase.  Since the average
electric field inside a dipolar fluid is zero, it is evident that
the mean-field contribution to the free energy also vanishes,
and all the non-trivial thermodynamics is, once again, the
result of electrostatic correlations.  

The thermodynamics 
of $DHS$ is particularly
tricky because of the unscreened long range interactions.  In fact
the very existence of the thermodynamic limit for $DHS$ has been 
proven only recently~\cite{BaGrWi98}.  Nevertheless, it has been taken for
granted that if the temperature is sufficiently low, the $DHS$ will
phase separate into coexisting liquid and gas phases.  Indeed,
all the theories 
have been predicting exactly this kind of behavior~\cite{GePi70,RuStHo73}.
It came, therefore, as a great surprise when the simulations in the early
$90's$ failed to locate the 
anticipated liquid-gas critical point~\cite{WeLe93,Ca93,LeSm93}.
Instead as the temperature was lowered, 
the simulations found chains of aligned dipoles. 
Formation of weakly interacting chains,  a consequence of strong
positional and directional correlations between the dipolar particles,
prevented the liquid-gas phase separation from taking 
place~\cite{Le99,Se96,Ro96,TaTeOs97}.

Electrostatic correlations are also
crucial in charged colloidal suspensions~\cite{RoHa97,HaLo00}. In
these systems the correlations come on two different levels.
First, there are very strong positional correlations between the
highly charged colloidal particles and their counterions. These 
correlations lead to charge renormalization~\cite{AlChGr84} and to screening
of Coulomb interactions between the colloidal particles. 
In water with monovalent
counterions, charge renormalization stabilizes 
colloidal suspension against phase 
separation~\cite{LeBaTa98,DiBaLe01}.
In the presence of multivalent counterions, however, the layers
of condensed counterions on different 
colloids can become strongly correlated, leading
to a net attraction between the like-charged colloids~\cite{Pa80,GuJoWe84} and
to the phase separation~\cite{LiLo99,Lo00}.  

A similar kind of behavior was also observed in a number of important   
biological systems.
Thus, it was noted that like-charged macromolecules can
attract each other in solutions containing multivalent counterions.
This attraction manifests itself in {\it in vitro} 
formation of toroidal aggregates of concentrated DNA~\cite{Bl91,Bl97}, 
similar to the
one found in  bacteriophage heads~\cite{Kl67}, and in  bundle
formation  of  F-actin and tobacco mosaic virus~\cite{Ta96}. 
A number of models have been suggested to explain these curious
phenomena.  The fundamental ingredient in all of these models
are the electrostatic correlations~\cite{HaLi97,ArStLe99,Sh99}.


Although the phenomena described above are quite complex,
we can get a long way towards understanding them by 
considering some surprisingly simple models and theories. 
In fact,
we shall demonstrate that a lot of the  physics of electrostatic
correlations is contained within the
Debye-H\"uckel theory~\cite{DeHu23} introduced $80$ years ago as a way 
of accounting for the unusual thermodynamic properties 
of strong electrolytes.

\begin{figure}  
\begin{center}
\includegraphics[width=8cm]{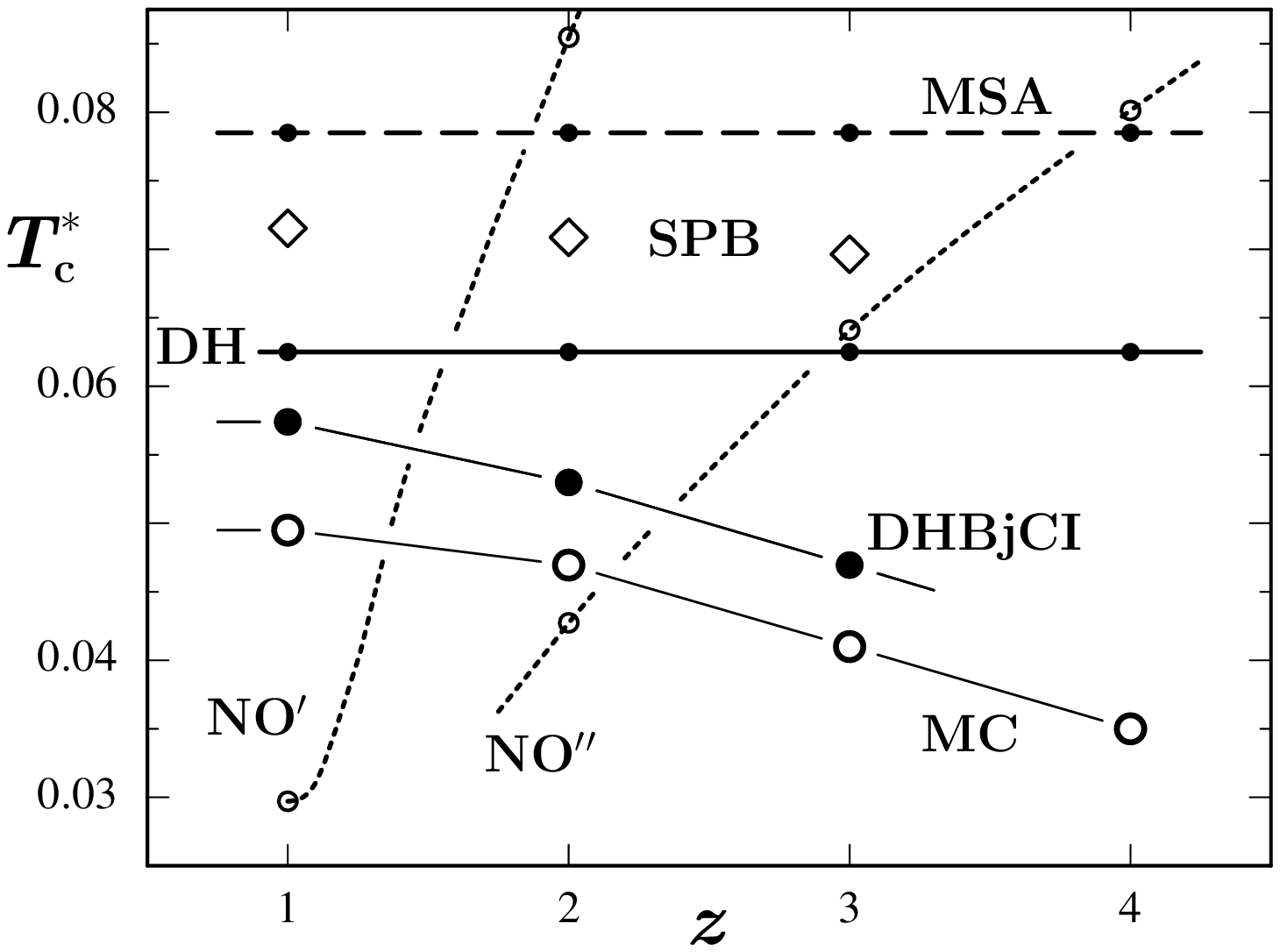}
\includegraphics[width=8cm]{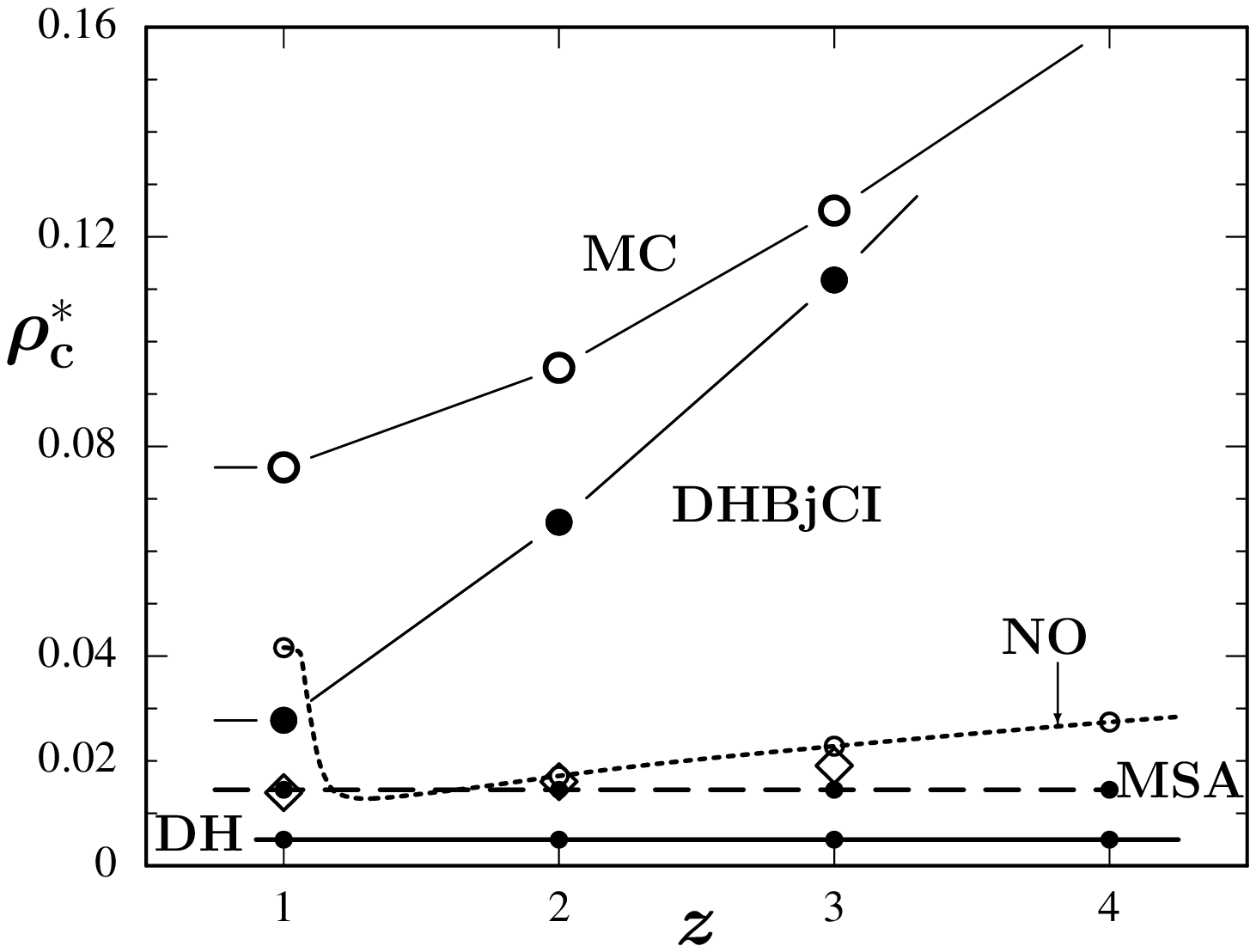}
\end{center}
\caption{Estimates of the reduced critical temperature, $T_c^*$, and
density, $\rho_c^*$, for the $(z:1)$, charge-asymmetric (but equisized)
primitive model showing, as labeled, the predictions of pure
Debye-H\"uckel theory (without hard cores), (b) the MSA, (c) the SPB
approximation~\cite{SaBhOu98}, (d) the Netz-Orland field-theoretic
treatment~\cite{NeOr99} which, for $T_c^*$,
have been divided by factors of $6$ and $12$ in order to bring them on
to the plot (see labels NO$^\prime$ and NO$^\prime$$^\prime$
respectively), and (e) the DHBjCI theory (following the Fisher-Levin
approach~\cite{FiLe93,LeFi96}), point $z=3$ is a preliminary result. The large
open circles for $z=2$ and $3$ represent, the Monte Carlo simulations
of Panagiotopoulos and Fisher~\cite{PaFi02},  
while the estimate for $T_c^*(4)$ 
follows from Camp and Patey~\cite{CaPa99}. (After S. Banerjee and
M. E. Fisher, to be published).}
\label{Fig0}
\end{figure}

Consideration of only simple physical  
theories in this review is partially
pedagogical, designed for a  broad audience
not necessarily familiar with the complex machinery
of correlation functions and field theories of the modern 
statistical mechanics.
For Coulomb systems there is, however, an additional
benefit. 
It is often found that the more sophisticated theories  
fail when
applied to  strongly correlated
charged fluids.  For example, the field theoretic
calculations of Netz and Orland~\cite{NeOr99} find that 
for charge asymmetric $(z:1)$ electrolytes the reduced
critical temperature is a strongly increasing 
function of charge
asymmetry.   
A completely  opposite behavior 
is observed in  computer
simulations, the critical temperature decreases and the critical density
increases with the charge asymmetry. In fact, the field theoretic
predictions for the critical temperature of asymmetric electrolytes 
are so far-off that they have to be divided by a factor
of six just to make them fit on the same graph (Fig. \ref{Fig0}) 
with the results of simulations
and of concurrent theories! The dramatic failure of field
theoretic calculations can be attributed to their  
intrinsically perturbative nature.
Similarly the integral equations, which
have proven to be  very successful for  simple molecular
fluids, fail to even converge for strongly 
asymmetric electrolytes.  Furthermore, it has  been
known for a long time that the 
Hypernetted Chain Equation (HNC)  which is often used 
to study the Coulomb systems~\cite{Be86a}, does not posses a true critical
region, but only a ``no solution zone'' on the border
of which  compressibility goes to zero with a square root
singularity~\cite{FiFi81,Be93}. This is a completely 
wrong behavior, since
the compressibility must diverge at the critical
point.   All these should be 
contrasted with the physically based Debye-H\"uckel-like theories,
which are in qualitative and often in  quantitative
agreement with the simulations and experiments, Fig. \ref{Fig0}.  
For some important
many body systems, however, 
the integral equations provide the most accurate
results.  For example,  predictions for the electrostatic free 
energy of the one component plasma obtained using 
the $HNC$ equation are in excellent agreement with the 
Monte Carlo simulations~\cite{Ng74}. The integral
equations were also the first to account for the correlation
induced attraction between the like-charged 
macromolecules~\cite{Pa80,KjMa84,KjMa86}.
In general, as long as one stays away from the phase transitions,
the integral equations provide one of the sharpest
tools available to a statistical physicist or chemist.  
Unfortunately,
the approximations involved in constructing the integral
equations are not very clear.  There exists a great
variety of closures to the Ornstein-Zernike equation,
each working  well for specific kind of problems.
Because of their complexity, I will not talk about 
the integral equations in this review, 
referring the interested reader to the literature~\cite{HaMc76}.

\section{Electrolyte solutions}   

Since the work of Faraday early in the $19'th$ century 
the flow of electricity has been associated
with the movement of charged particles. The nature of ions 
(from Greek meaning wanderers), however, was not established.  
It was Arrhenius who in $1887$, following the
experimental work of van't Hoff on osmotic pressure of
electrolyte solutions, proposed that when salts and acids
are dissolved in water they become ionized~\cite{Ar87}. 
Arrhenius suggested
that $NaCl$ dissociates forming cations $Na^+$ and
anions $Cl^-$.  
In the spirit of the mean-field theory introduced earlier
by van der Waals~\cite{Wa73}, Arrhenius argued that since
the anions and the cations are on average 
uniformly distributed throughout the
solution,
the average electrostatic potential inside the
electrolyte is zero.  He then concluded that on average there
should not be any interaction between the ions, and the osmotic
pressure of, say, $1M$ solution of $NaCl$  
should be equivalent
to the osmotic pressure of $2M$ solution of a non-electrolyte.
All the deviations from this simple rule Arrhenius attributed to
the incomplete dissociation of electrolyte, which he then treated
as a problem of chemical equilibrium, the thermodynamics of 
which has already been developed by Gibbs
ten years earlier~\cite{Gl46}.  
Soon, however, it became clear that while the theory was
working  well for weak electrolytes, such as Br\o nsted acids
and bases, it failed for strong electrolytes 
such as $NaCl$ and $HCl$, which remained
fully ionized even at fairly large concentrations.
The disagreements between the theory and the experiment could not be
accounted for by the postulate of 
incomplete dissociation. It appeared that there
was a fundamental flaw
in the theory advanced by Arrhenius, which
relied on the mean-field assumption of  non-interacting ions.
The situation remained unclear
for about 30 years, with various proposals made on how to 
incorporate the ionic interactions into the framework of Arrhenius'
theory~\cite{Gl46}.  None of these have proven very successful
at explaining the experimental measurements, 
until Debye and H\"uckel published their, now
famous, theory of strong electrolytes~\cite{DeHu23}. The fundamental insight
of Debye and H\"uckel $(DH)$ was to realize that although ions {\it  on
average} are randomly distributed, there exist strong positional
correlations between the anions and cations. 
The depth of Debye's insight can be judged from the fact that
he understood the role of electrostatic 
correlations significantly before the correlation functions
became the standard tool of working physicists.
Since so many of our results will be based on the
fundamental ideas of Debye and H\"uckel, their
theory will provide the starting point for our
discussion of thermodynamics of electrostatic correlations.

\subsection{The Debye-H\"uckel theory}

Consider the  simplest model of an electrolyte solution confined to
volume $V$.  The
$N$ ions will be idealized as hard spheres of diameter $a$ carrying
charge $\pm q$ at their centers. The charge neutrality of solution
requires that $N_+=N_-=N/2$. The solvent will be modeled
as a continuum of dielectric constant $\epsilon$.
\begin{figure}  
\begin{center}
\includegraphics[width=6cm]{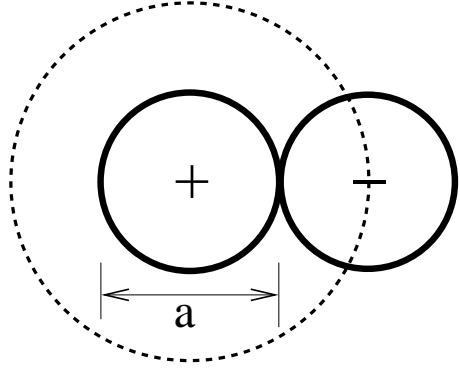}
\end{center}
\caption{The configuration of closest approach between two oppositely
charged ions. The dashed curve delimits the region into which no ions
can penetrate, due to the hard core repulsion.}
\label{Fig1}
\end{figure}
Although the average potential inside the electrolyte is zero,
there are strong positional correlations between the oppositely
charged ions.  It is convenient to work in spherical coordinates. 
To calculate the correlational contribution to the
Helmholtz free energy, let us fix one ion of charge $+q$ at the origin
$r=0$ and see how the other
ions will distribute around it, see Fig. \ref{Fig1}. 
Inside the region $0<r \le a$ there are no other charges except for the
one fixed at the origin, and 
the electrostatic potential $\phi(r)$ satisfies
the Laplace equation,
\begin{equation}
\label{0a}
\nabla^2 \phi=0 \;.  
\end{equation}
For $r>a$ the electrostatic potential satisfies the Poisson equation
\begin{equation}
\label{0}
\nabla^2 \phi=-\frac{4 \pi}{\epsilon} \rho_q(r) \;,
\end{equation}
where the charge density can be expressed in terms of the charge-charge
correlation functions $g_{++}(r)=g_{--}(r)$ and $g_{+-}(r)=g_{-+}(r)$ 
\begin{equation}
\label{1}
\rho_q(r)=q \rho_+ g_{++}(r)-q\rho_- g_{+-}(r) \;.
\end{equation}
The average densities of positive and negative ions are
$\rho_+=N_+/V$, $\rho_-=N_-/V$; $\rho_+=\rho_-=\rho/2$.

The correlation
functions can be written in terms of the 
potential of mean force $w_{ij}$   
\begin{equation}
\label{1a}
g_{ij}(r)=e^{-\beta w_{ij}(r)} \;, 
\end{equation}
where $\beta=1/k_BT$. The $w_{ij}(r)$ is the work required to bring
ions $ i$ and $ j$ from infinity to separation $r$ inside the electrolyte
solution.
In their paper Debye and H\"uckel made an implicit approximation of
replacing the potential of mean force by the electrostatic potential
\begin{equation}
\label{1b}
w_{ij}(r)=q_j \phi_i(r) \;, 
\end{equation}
where $q_j$ is the charge of $j'th$ ion and $\phi_i(r)$ is the electrostatic
potential at distance $r$ from the ion $i$ fixed at the origin $r=0$.
With this approximation, 
Eq.(\ref{0}) reduces to the non-linear Poisson-Boltzmann equation $(PB)$,
\begin{equation}
\label{2}
\nabla^2 \phi=-\frac{4 \pi}{\epsilon} 
\left[q \rho_+ e^{-\beta q \phi} - q \rho_- e^{+\beta q \phi}\right]=
\frac{4 \pi \rho q}{\epsilon}\sinh(\beta q \phi) \;.
\end{equation}
Debye and H\"uckel proceeded
to linearize this equation.  Technically, linearization is only
valid if $\beta q \phi \ll 1$, however, being practically minded Debye and 
H\"uckel
linearized first and worried about the consequences later. 
As was noted later by Onsager, linearization of Eq.(\ref{2}) is a 
necessary step in order to produce a self-consistent theory\cite{On33}. 
The linearized Poisson-Boltzmann
equation reduces to the Helmholtz equation
\begin{equation}
\label{3}
\nabla^2 \phi=\kappa^2 \phi\;,
\end{equation}
where the inverse Debye length is
\begin{equation}
\label{3a}
\xi_D^{-1}\equiv \kappa=\sqrt{\frac{4 \pi q^2 \rho}{k_B T \epsilon}}.
\end{equation}

The Laplace equation (\ref{0a}) for $r\le a$ 
and the Helmholtz equation (\ref{3}) for $r>a$
must be integrated, subject to the boundary condition of 
continuity of the electrostatic potential and the
electric field across the boundary surface $r=a$.
For $ r \le a$ the electrostatic potential is found to be
\begin{eqnarray}
\label{4}
\phi_<(r)=\frac{q}{\epsilon r}-\frac{ q\kappa}{\epsilon (1+\kappa a)} \;, 
\end{eqnarray}
while for  $ r > a$,
\begin{eqnarray}
\label{5}
\phi_>(r)=\frac{q \theta(\kappa a) e^{-\kappa r}}{\epsilon r} \;, 
\;\;\;\; 
\theta(x)=\frac{e^{x}}{(1+x)}\;.
\end{eqnarray}
Equation (\ref{5}) shows that the electrostatic potential produced by
the central charge is exponentially screened 
by the surrounding ionic cloud.  Because
 of the hardcore repulsion the 
screening, however, appears only at distances larger
than $r=a$. This accounts for the presence of  geometric factor 
$\theta(\kappa a)$ in Eq.~(\ref{5}).
The screening of electrostatic interactions inside the electrolyte
solutions and plasmas is responsible for the existence of thermodynamic
limit in these systems with extremely long range forces.
 
The electrostatic potential $\phi_<(r)$, Eq.~(\ref{4}), consists of
two terms: the potential produced by the central ion $q/\epsilon r$,
and 
the electrostatic potential induced by the surrounding ionic cloud,
\begin{equation}
\label{5a}
\psi= -\frac{q\kappa}{\epsilon (1+\kappa a)}. 
\end{equation}
The electrostatic free energy can now be obtained
using the Debye charging process in 
which all the ions are simultaneously charged
from zero to their full charge,
\begin{equation}
\label{6}
F^{el}=N q \int_0^1 d \lambda \psi(\lambda q)\;.
\end{equation}
The calculation is very similar to the one used to
obtain the electrostatic energy stored in a 
capacitor.
While performing the charging it is important to remember 
that $\kappa(\lambda q)=\lambda \kappa(q) $. 
Defining
the free energy density as 
$f=F/V$,  
the integral in Eq.~(\ref{6})
can be performed explicitly yielding
\begin{equation}
\label{7}
\beta f^{el}=\frac{\beta F^{el}}{V}=-\frac{1}{4 \pi a^3}\left[\ln(\kappa a+1)-\kappa a+\frac{(\kappa a)^2}{2}\right]\;.
\end{equation}
For large dilutions Eq.~(\ref{7}) reduces to the famous Debye limiting law, 
\begin{equation}
\label{7a}
\beta f^{el} \approx -\frac{\kappa^3}{12 \pi} \sim -\left(\frac{\rho}{T}\right)^{3/2} \;.
\end{equation}
Given the free energy, the limiting laws for the osmotic pressure 
and activity can be easily found~\cite{LeFi96}.

The free energy is not analytic at $\rho=0$.   
The singularity at $\rho=0$ is a consequence of long-range 
Coulomb interactions,
which also manifest themselves in the divergence of the standard
virial expansion~\cite{FiLiLe95}.  
The total free energy of the electrolyte $F$ 
is the sum of electrostatic Eq.~(\ref{7}), and 
entropic contributions.   The entropic contribution to the
free energy arises from the integration 
over the momentum degrees of freedom in the partition function,
and is equivalent to the free energy of an ideal gas,  
\begin{equation}
\label{8}
 \beta F^{ent}=N_+\ln[\rho_+ \Lambda^3]-N_+ +
N_-\ln[\rho_- \Lambda^3]-N_-=N\ln[\rho \Lambda^3/2]-
N\;,
\end{equation}
where the de Broglie thermal wavelength is
\begin{equation}
\label{8aa}
\Lambda=\frac{h}{\sqrt{2 \pi m k_B T} }\;.
\end{equation}
The osmotic pressure of the electrolyte is 
\begin{equation}
\label{8a}
P=-\frac{\partial F}{\partial V} \Big\arrowvert_{N}\;,
\end{equation}
which can also be expressed in terms of the  Legendre transform of the
negative free energy density $-f$ \cite{LeFi96},
\begin{equation}
\label{9}
 P=-f+ \mu \rho \;,
\end{equation}
where the chemical potential is 
\begin{equation}
\label{9a}
\mu= \frac{\partial F}{\partial N}\Big\arrowvert_{V}=
\frac{\partial f}{\partial \rho} \;.
\end{equation}

It is a simple matter to see 
that below the critical temperature $T_c$ the total free energy 
$F=F^{ent}+F^{el}$
fails to be a convex function of the electrolyte concentration. 
This implies the presence of a phase transition.  Alternatively
the phase separation can be observed 
from the appearance of a  van der Waals loop in the osmotic 
pressure Eq.~(\ref{9}), below the critical temperature $T_c$.  
The critical parameters  are determined from
\begin{equation}
\label{9aa}
\frac{\partial P}{\partial \rho}=0 \;,
\end{equation}
\begin{equation}
\label{9ab}
\frac{\partial^2 P}{\partial \rho^2}=0 \;.
\end{equation}
The coexistence
curve can be obtained using the standard Maxwell construction.  
It is convenient
to define the reduced temperature and density as
$T^*=k_B T a \epsilon/q^2$ and $\rho^*=\rho a^3$.  
The critical point of the plasma, within the $DH$ theory, is found 
to be located at~\cite{FiLe93,LeFi96}
\begin{equation}
\label{9b}
T^*_c=\frac{1}{16}\;,
\end{equation}
and 
\begin{equation}
\label{9c}
\rho_{c}^*=\frac{1}{64 \pi} \;.
\end{equation}
It is interesting to note
that at criticality $\kappa = 1/a$.  This means that in spite of a
very low concentration of 
electrolyte at the critical point, the screening
remains very strong.  
We also observe that the reduced critical
temperature for the electrolyte is almost an order of magnitude lower,
than for systems in which the particles interact by the short-ranged 
isotropic potentials.  
Since the critical point within the
$DH$ theory occurs at 
extremely low density, we are justified in neglecting
the excluded volume contribution to the total free energy.

Phase separation of an electrolyte or of a two component
plasma is the result of an electrostatic instability arising from the
strong positional correlations between the oppositely charged ions.  
This mechanism is very different from 
the one driving  the phase separation in
systems dominated by the  short ranged isotropic forces.  In that case
the thermodynamic instability is a consequence of the  
competition between the interparticle attraction  
and the hardcore repulsion.

The reduced temperature can  be written as $T^*=a/\lambda_B$, where
$\lambda_B=q^2/k_B T\epsilon$ is the Bjerrum length.  
For water at room temperature
$\lambda_B \approx 7$ \AA.  This means that one would need ions
of size less than $0.4$ \AA,  
in order to observe phase separation at
room temperature.  This is clearly impossible since the minimum 
hydrated ionic size is about $2-4$ \AA.  Therefore, in order to see 
phase separation, one is required to look for solutions with
$\lambda_B$ on the order of $40$ \AA $\,$ or more. For water such large
values of $\lambda_B$ correspond to temperatures 
well below the freezing.  An alternative is to work with
organic solvents which have dielectric constants significantly
lower than water. This was the strategy adopted by K.S. Pitzer in
his studies of ionic criticality~\cite{Pi95,NaPi95,NaPi94}.  
Pitzer used liquid salt
triethyl-n-hexylammonium triethyl-n-hexylboride $(N_{2226}B_{2226})$
in the diphenyl ether.  With this he was able to observe the critical point
at room temperature. Pitzer's work has provoked a lot of stimulating 
controversy because his measurements suggested that the Coulombic
criticality belonged to a new universality class~\cite{Wi98}.  At first sight this
might not seem very surprising, after all the Coulomb force is 
extremely long ranged.  On further reflection the situation is
not so clear. Although the bare interaction potential 
between any two ions is long ranged, 
inside the electrolyte solution it is screened by the surrounding
particles, as is seen from Eq.~(\ref{5}). The effective interaction
potential, therefore, is short ranged, which should place the
ionic criticality firmly in the Ising universality class.  In fact
all the theoretical arguments lead to this conclusion, which
seems to be contradicted by the Pitzer's experiments.  In principle,
it is possible that one has to be very close to the critical point
before the Ising behavior becomes apparent.  However, even this
conclusion is hard to justify theoretically.  Estimates of the
Ginzburg criterion suggest that the width of the critical region
for the Coulombic criticality should be comparable to that of systems
with short ranged isotropic interactions~\cite{FiLe93,FiLe96}. 
The situation remains unclear.

An alternative to working with electrolyte solutions is to
study molten salts, which are classical two component plasmas.
In this case the dielectric constant can be
taken to be that of vacuum, and  ions are no longer hydrated.
The reduced critical temperature $T^*=1/16$ and the characteristic
ionic diameter of about $2$ \AA, imply that at criticality 
$\lambda_B \approx 30$, which means that the critical
point for a molten salt is located at about $5000K$.  It is,
indeed, very hard to study critical phenomena at such high
temperatures!  It seems, therefore, that we are stuck with the low dielectric 
solvents. An alternative is the computer simulations, which are becoming
sufficiently accurate to allow  measurements of the critical
exponents, at least for symmetric 1:1 electrolytes. Indeed the most recent
simulations suggest that the Coulombic criticality 
belongs to the Ising universality class~\cite{LuFiPa01}.  

\subsection{The Bjerrum association}
\label{Bjerrum}
The $DH$ theory presented in the previous section was based on the
linearization of the Poisson-Boltzmann equation.  In view of 
the strong
screening and the rapid decrease of the electrostatic potential away
from the central ion, such a linearization can be justified at 
intermediate
and long distances. It is clear, however, that the linearization 
strongly diminishes 
the weight of configurations in which two oppositely
charged ions are in a close proximity. Linearization underestimates
the strength of electrostatic correlations which result in dipole-like
structures.  At low reduced temperatures 
characteristic of the critical point, these configurations should be
quite important and must be taken into account.  One way of doing 
this, while preserving the
linearity of the theory, is to postulate the existence of 
dipoles with concentration governed by the
law of mass action.  In the leading-order
approximation the dipoles can be treated 
as ideal non-interacting specie~\cite{Bj26,Eb68,FaEb71}. 
The total number of particles $N=\rho V$ is then subdivided 
into monopoles $N_1=\rho_1 V$ and  
dipoles $N_2=\rho_2 V$.  The  particle conservation requires that, 
$N=N_1+2 N_2$.  The free energy  of the mixture is 
$F=F_1^{ent}+F_2^{ent}+F^{el}$, where $F^{el}$ and 
$F_1^{ent}$ are the entropic and the electrostatic free energies 
of the monopoles, given by the Eqs.~(\ref{7}) and (\ref{8}),
but with $N \rightarrow N_1$ and  $\rho \rightarrow \rho_1$.
The entropic free energy of dipoles is,
\begin{equation}
\label{10}
 \beta F^{ent}_2=N_2\ln[\rho_2 \Lambda^6/\zeta_2]-N_2 \;,
\end{equation}
where the internal partition function of a dipole is,
\begin{equation}
\label{11}
\zeta_2(R)=4 \pi \int_a^R r^2 dr \,\exp\left(\frac {\beta q^2}{\epsilon r}\right) \;.
\end{equation}
At low temperatures, the  precise value of the cutoff $R$ at which the
two ions can be considered to be  associated is not very
important.  Following the original suggestion of Bjerrum\cite{Bj26} we
can take this value to be the inflection point of the integral
in Eq.~(\ref{11}), $R_{Bj}= \lambda_B/2$.  This choice corresponds
to the minimum of integrand in Eq.~(\ref{11}), which
in turn can be interpreted as the probability of finding two 
oppositely charged
ions at the separation $r$. The minimum then correspond to 
a liminal between bound and 
unbound configurations.  A much more careful analysis
of the dipolar partition function has been carried out
by Falkenhagen and Ebeling based on the 
resummed virial expansion~\cite{FaEb71}.  They found that
that the low temperature expansion of 
the Bjerrum equilibrium
constant is identical to the equilibrium constant which can
be constructed on the basis of the 
resummed virial expansion. Since we are interested
in the low temperature regime where the critical
point is located, the Bjerrum equilibrium constant, 
$\zeta_2 \equiv \zeta_2(R_{Bj})$, will be sufficient.

It is important to keep in mind that at this level
of approximation the electrostatic free energy $F_1^{ent}$ 
is only a function of the density of free unassociated ions $\rho_1$, since
the dipoles are treated as ideal non-interacting specie.
The concentration of dipoles is obtained from the law of mass action 
\begin{equation}
\label{11a}
 \mu_2=\mu_++\mu_- \;,  
\end{equation}
where the chemical potential of a specie $s$ is  
\begin{equation}
\label{11b}
\mu_s=\frac{\partial F}{\partial N_s}\Big\arrowvert_{V} \;.  
\end{equation}
Substituting the expression for the 
total free energy into the law of mass
action leads to
\begin{equation}
\label{12}
\rho_2= \frac{1}{4} \rho_1^2 \zeta_2 \,e^{2\beta \mu^{ex}}\;,
\end{equation}
where the excess chemical potential is   
$\mu^{ex}=\partial f^{el}/\partial \rho_1$. 
The critical point can be located from the study of the
convexity of the total free energy
as a function of ion concentration $\rho$.  
There is, however, a simpler way~\cite{LeFi96}.
We observe that at Bjerrum level of approximation, dipoles are 
ideal non-interacting specie.  This means that they are only present
as spectators and do not interact with the monopoles in any way. 
This implies that
only the monopoles can drive the phase separation.  Thus, 
at the critical point the temperature must still be $T^*_c=1/16$ and
the density of monopoles must still remain $\rho_{1c}^*=1/64 \pi$,
as in the case of the pure $DH$ theory.  
The corresponding density of dipoles at criticality is then given
by Eq.~(\ref{12}), with $T^*_c=1/16= 0.0625$ and 
$\rho_{1c}^*=1/64 \pi=0.00497$. 
We find that at the critical point 
the density of dipoles is $\rho_{2c}^* \approx 0.02$. 
In the vicinity of the critical point 
there are many more dipoles than monopoles,
$\rho_{2c}^*/\rho_{1c}^*\approx 4$. Within the Bjerrum approximation
the non-linear correlations, in the form of dipoles,  
do not affect the critical temperature, but
strongly modify the critical density,  $\rho_c^*
=\rho^*_{1c}+2\rho^*_{2c}=0.045$.  
In spite of the crudeness of approximations,
the location of the critical point agrees reasonably well with the
Monte Carlo simulations~\cite{CaLeWe97,OrPa99,YaPa99}, $T^*_c=0.051$ and $\rho_c^*=0.079$.  
The coexistence curve, however, is found to have
an unrealistic ``banana'' shape~\cite{LeFi96}.  
To correct this deficiency one must
go beyond the ``ideal'' dipole approximation and 
allow for the dipole-ion
interaction~\cite{FiLe93,LeFi96}.  
Most of the fundamental physics of electrostatic 
correlations, however, is already captured at the level of the
Bjerrum approximation.

\section{Two-dimensional plasma and the Kosterlitz-Thouless transition}
\label{KT}

An electrostatic system which over the years has attracted much attention
is the two-dimensional plasma of positive and negative ions 
interacting by
a logarithmic potential.  The great interest 
in  2d plasma is due to the fact that 
various important physical systems
can be mapped directly onto it.  Examples include superfluid $^4$He
films, two-dimensional crystalline solids, and $XY$ magnets~\cite{Ne83}. 
Although a continuous symmetry can not be broken in 
two dimensions~\cite{Me68},
if the Hamiltonian of a system is invariant under an Abelian group,
a finite temperature phase transition
is possible.  This transition occurs as the result of 
unbinding of the topological defects or ``charges''.
The defect-mediated phase transitions belong to
the universality class of a two-dimensional plasma.

Thirty years ago Kosterlitz and Thouless $(KT)$ have
presented a renormalization group study of the 2d
plasma~\cite{KoTh73}.  
They concluded
that at sufficiently low temperature, the 2d plasma becomes
an insulator.  All the positive and negative ions pair-up 
forming dipoles. The
metal-insulator transition was found to be of an infinite order, 
characterized by an essential singularities in the thermodynamic functions.
The KT analysis, however, was restricted to the low ionic densities
and it is not clear what happens when the concentration of charged 
particles is increased.  It is tempting to apply to the 2d plasmas  
analysis similar to the one 
presented earlier for the 3d electrolytes~\cite{LeLiFi94}.

We shall, then, study a fluid  of disks with
diameter $a$ and charge $\pm q$. The solvent is a uniform medium
of dielectric constant $\epsilon$. The bare interaction potential 
for two ions $(i,j)$ separated by distance $r$ is 
\begin{equation}
\label{12a}
\varphi(r)=-\frac{q_i q_j}{\epsilon} \ln(r/a)\;.
\end{equation}
As in the case of 3d electrolyte, 
the mean-field contribution to the electrostatic 
free energy is zero, and all the important physics 
comes from the electrostatic correlations. We shall account
separately for the long-ranged and the 
short-ranged correlations.  The short
ranged correlations lead to the formation of dipolar pairs of
density $\rho_2$, while the long ranged correlations produce the
screening.  As in the case of 3d electrolyte, the total density of hard
discs is divided between the dipoles and the monopoles, 
so that $\rho=\rho_1+2 \rho_2$. 

To calculate the electrostatic
free energy we fix one ion and study the distribution of other
particles around it. It is important to recall that the  
Poisson equation in 2d,
\begin{equation}
\label{13}
\nabla^2 \phi(r)= -\frac{2 \pi}{\epsilon} \rho_q(r)\;,
\end{equation}
differs from the
one in 3d by the normalization factor~\cite{LeLiFi94},
$2 \pi$ has replaced the $4 \pi$ of the 3d Poisson equation.  
  
As before, we shall approximate
the potential of mean force by the electrostatic potential and
then linearize the Boltzmann factor.  Linearization is 
compensated by the allowance for 
dipolar formation.

\begin{figure}  
\begin{center}
\includegraphics[width=6cm]{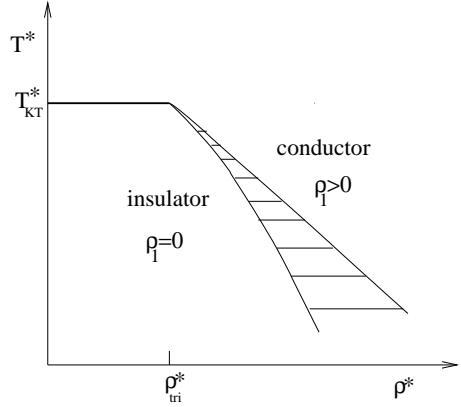}
\end{center}
\caption{Phase diagram for the two dimensional plasma within
the Debye-H\"uckel-Bjerrum approximation. 
We expect the fluctuations to renormalize the 
Kosterlitz-Thouless line, shifting it from its horizontal position
and making it density dependent.  The topology of the phase diagram
should, however, remain the same.}
\label{Fig2}
\end{figure}

The electrostatic potential for distances
$r \le a$  satisfies the Laplace equation $\nabla^2 \phi=0$, while
for $r>a$  the potential satisfies the 
Helmholtz equation $\nabla^2 \phi= 
\kappa^2 \phi$, with 
\begin{equation}
\label{14}
\kappa=\sqrt{\frac{2 \pi q^2 \rho_1}{k_B T \epsilon}} \;,
\end{equation}
This equations
can be easily integrated yielding the electrostatic potential,
\begin{equation}
\label{14a}
\phi_<(r)=-\frac{q}{\epsilon}\ln(r/a)+
\frac{q}{\epsilon}\frac{ K_0(\kappa a)}{\kappa a K_1(\kappa a)} 
\;\;\; for \;\;\; r \le a\;,
\end{equation}
and 
\begin{equation}
\label{14b}
\phi_>(r)=\
\frac{q}{\epsilon}\frac{ K_0(\kappa r)}{\kappa a K_1(\kappa a)} 
\;\;\; for \;\;\; r > a \;,
\end{equation}
where $K_\nu(x)$ are the modified Bessel functions of order $\nu$.
For large distances, the electrostatic potential 
decays exponentially fast.  
Just as in three dimensions, 
the electrostatic interactions are screened inside the
two dimensional plasma,
\begin{equation}
\label{14c}
\lim_{r \rightarrow \infty}\phi_>(r) \approx 
\frac{q e^{-\kappa r}}{\epsilon \kappa a K_1(\kappa a)}\sqrt{\frac{\pi}{2 \kappa r}}  \;.
\end{equation}
Eq.~(\ref{14a}) consists of two terms, the potential
produced by the fixed ion and the induced potential due to
ionic atmosphere,
\begin{equation}
\label{14d}
\psi= \frac{q}{\epsilon}\frac{ K_0(\kappa a)}{\kappa a K_1(\kappa a)} \;.
\end{equation}
Given the induced  potential, the electrostatic free energy can
be obtained using the familiar Debye charging process, Eq.~(\ref{6}).
We find the electrostatic free energy density of a 2d plasma
to be
\begin{equation}
\label{15}
\beta f^{el}= \frac{1}{2 \pi a^2} \ln[\kappa a K_1(\kappa a)]\;.
\end{equation}
At the Bjerrum level of approximation, the electrostatic free
energy depends only on the density of monopoles. 
The total free energy density
is  the $f=f_1^{ent}+f_2^{ent}+f^{el}$, where
\begin{equation}
\label{16}
\beta f^{ent}_1=\rho_1\ln[\rho_1 \Lambda^2/2]-\rho_1 \;
\end{equation}
and
\begin{equation}
\label{17}
\beta f^{ent}_2=\rho_2\ln[\rho_2 \Lambda^4/\zeta_2]-\rho_2 \;.
\end{equation}
The internal partition function for a 2d dipole is
\begin{equation}
\label{18}
\zeta_2(R)=2 \pi \int_a^R r dr \,\exp\left[-\frac{\beta q^2}{\epsilon} 
\ln\left(\frac{r}{a}\right)\right] \;.
\end{equation}
It is convenient to define the reduced temperature and density
as $T^*=k_B T \epsilon/q^2$ and $\rho^*=\rho a^2$.  We note that
for low temperatures, $T^*<1/2$, the integral in  Eq~(\ref{18})
converges uniformly
as $R \rightarrow \infty$.  In this regime it is possible, therefore, to 
define
the internal partition function of dipole as 
\begin{equation}
\label{19}
\zeta_2 \equiv \zeta_2(\infty)=\frac{2 \pi a^2 T^*}{1-2 T^*}\;.
\end{equation}
The thermodynamic equilibrium requires that for fixed 
volume and number of particles  the   
Helmholtz free energy  be minimum.
This is equivalent to the law of mass action Eq.~(\ref{11a}), which upon
the substitution of free energy simplifies to Eq.~(\ref{12}).
In the limit of small concentrations, the excess chemical potential  
can be expanded in powers of $\rho_1$ yielding
\begin{equation}
\label{20}
\beta \mu^{ex}=-\frac{1}{2 T^*}[\gamma_E+\ln(\kappa a/2)]\;,
\end{equation}
where $\gamma_E$ is the Euler constant. 
Substituting Eq.~(\ref{20}) into Eq.~(\ref{12}), 
we find that the concentration of dipoles 
in the limit $\rho \rightarrow 0$ scales as 
\begin{equation}
\label{20a}
\rho_2 \sim \rho_1^{\theta(T^*)}\;,
\end{equation}
where 
\begin{equation}
\label{20b}
\theta(T^*)=2-\frac{1}{2T^*}\;.
\end{equation}
For $T^*<1/4$, the exponent $\theta(T^*)<0$, and in the limit 
$\rho_1 \rightarrow 0$ the law of mass action can not
be satisfied.  This
means that in the temperature density plane $(T^*,\rho^*)$, 
for sufficiently
small densities, the line
$T^*=1/4$ corresponds to the critical locus of metal-insulator 
transitions.  
Below this line, and for sufficiently small ionic concentrations, 
no free monopoles can exist.
All the ions  are paired up into neutral dipolar pairs. The 
critical line terminates at the tricritical point located at
$T^*_{KT}=1/4$ and 
\begin{equation}
\label{20c}
\rho^*_{tri}=\frac{e^{-4 \gamma_E}}{8 \pi} \simeq 0.003954\;.
\end{equation}
For $T^*<1/4$ and $\rho^*>\rho^*_{tri}$ there is a phase separation
between an insulating vapor and a conducting liquid phases, 
Fig \ref{Fig2}. 
As the critical line is approached from  high temperatures, the Debye
length diverges as 
\begin{equation}
\label{20d}
\xi_D \equiv \kappa^{-1} \sim e^{c(\rho)/t^\nu}\;,
\end{equation}
where 
\begin{equation}
\label{20e}
c(\rho)=\frac{1}{4}\ln \left(\frac{\rho_{tri}}{\rho}\right)\;,
\end{equation}
and 
\begin{equation}
\label{20ee}
t=\frac{T-T_{KT}}{T}\;.
\end{equation}
The critical exponent is $\nu=1$.
The Kosterlitz-Thouless (KT) renormalization group 
theory~\cite{KoTh73} predict the same 
behavior for $\xi_D$ except that $\nu=1/2$. The $KT$ theory, however,
leaves unanswered the question of  what happens to the 
metal-insulator transition for higher plasma concentrations.  
The  current theory, on the other hand,  
shows that the critical line terminates in a
tricritical point, after which the metal-insulator transition becomes 
first order~\cite{LeLiFi94}. 
This topology is also consistent with the findings of 
Monte Carlo simulations~\cite{Ca94,OrPa96}. 
A more sophisticated theory introduced by Minhagen~\cite{Mi87},
leads to a very similar phase diagram, except that the tricritical point 
is replaced by a critical end-point.

We see that the electrostatic correlations 
are even more important in 2d than in 3d. 
While in three dimensions the electrolyte phase separates
into the coexisting liquid and gas phases, both of which 
contain  monopoles and
dipoles,  in  two dimensions 
the low density vapor phase does not contain
any free charges and is an insulator.

\section{The one component plasma}
\label{OCP}

The one component plasma $(OCP)$ is probably the 
simplest model of a Coulomb system.  
It consists of point ions, all of the same sign, 
inside a rigid
neutralizing background of opposite charge. In spite of
its simplicity the model is relevant for many
physical systems.  Some examples  are the interiors of stars,
liquid metals, and magnetically confined electrons. 
As an approximation, $OCP$ is
particularly important since it provides a framework in which
ionic correlations can be calculated. Thus, the electrostatic free
energy of a homogeneous $OCP$ can be used
to account for the counterion correlations 
in colloidal suspensions and polyelectrolyte solutions.  
The $OCP$ has been extensively studied over the years.  A review  
of the subject, which still remains very actual,  has been 
presented by Baus and Hansen some twenty years ago~\cite{BaHa80}. 

In the spirit of the current work we shall, however, confine
our attention to simple analytical 
theories of the $OCP$~\cite{No84,TaLeBa99}.
Our model consists of $N$ ions each carrying charge $q$ inside a
uniform neutralizing background of dielectric constant $\epsilon$ 
and volume $V$. The average ionic density is $\rho=N/V$.
The mean electrostatic potential inside the $OCP$ is zero and the 
free energy is, once again, entirely due to 
positional correlations between
the ions.  The advantage of working with the $OCP$ is that 
it contains only two independent length scales: the average
separation between the particles $d=(4 \pi \rho/3)^{-1/3}$ and
the Bjerrum length $\lambda_B=q^2/k_B T\epsilon$. 
There is only one dimensionless parameter on which all the 
thermodynamic quantities depend, $\Gamma=\lambda_B/d$. 
This is quite distinct
from the electrolytes and two component plasmas,
for which besides $d$ and $\lambda_B$ there
is a third length scale  --- the ionic diameter $a$ --- 
and  the thermodynamics  is  
parametrized by two dimensionless quantities
$T^*$ and $\rho^*$.  For electrolytes the limit $a \rightarrow 0$ does not
exist, since for point particles 
the free energy can be lowered indefinitely by collapsing the system 
into point-like dipolar pairs.

The calculation of free energy of the $OCP$ proceeds 
along the lines of the Debye-H\"uckel
theory.  We fix one ion and study the distribution of other ions
around it.  The electrostatic potential satisfies the Poisson equation,
Eq.~(\ref{0}), with the charge density given by 
\begin{equation}
\label{20f}
\rho_q(r)=q \rho g(r)-q\rho \;.
\end{equation}
The correlation function can be expressed in terms of the potential
of mean force $w(r)$ as 
\begin{equation}
\label{20g}
g(r)=e^{-\beta w(r)} \;.  
\end{equation}
In the spirit of the Debye-H\"uckel theory we replace $w(r)$ by  
$q \phi(r)$.  This approximation entails
neglect of electrostatic correlations inside the ionic cloud which
surrounds the central ion.  The approximation should be quite
good as long as $\Gamma$ is not too large.  As the next step,
linearization of the Boltzmann factor leads to the charge density
of particularly simple form,
\begin{equation}
\label{21}
\rho_q(r)=- \frac{\epsilon \kappa^2}{4 \pi} \phi(r)\;,
\end{equation}
where $\kappa^2=4 \pi \beta q^2 \rho/\epsilon$.  
Substituting this into the Poisson equation we, once again, find
the familiar Helmholtz equation (\ref{3}).  
This can be easily integrated 
yielding the electrostatic potential of Yukawa form,
\begin{equation}
\label{22}
\phi(r)=\frac{q e^{-\kappa r}}{\epsilon r}\;.
\end{equation}

The charge density must be bounded from below 
$\rho_q \ge -q \rho$.  However, considering Eqs.~(\ref{21}) and
(\ref{22}), it is evident that this condition is  violated
for sufficiently small separations from the central ion.  
Something must have gone seriously
wrong.  It is easy to trace the problem to the linearization of the 
Boltzmann factor.  Clearly at short distances
linearization is not justified since the strong electrostatic
repulsion between the ions results in a very large electrostatic
energy. It appears that in order to understand the physics of the
$OCP$ one has to solve the full 
non-linear Poisson-Boltzmann equation, a task which cannot
be performed analytically.  
Fortunately not everything is lost.
A way out was suggested by Nordholm, who noted that the strong repulsion
between like-charged ions  results in an effective 
hole surrounding the central
ion~\cite{No84}.   Very few ions can 
penetrate inside the hole since this  costs
them too much electrostatic energy.  Inside this
correlation hole, $r\le h$, the electrostatic potential satisfies the
Poisson equation (\ref{0}) with a uniform background charge density, 
$\rho_q \approx -q \rho$.  In
the outside region, where the electrostatic interaction is much weaker,
the linearization of the Boltzmann factor is justified and the potential
satisfies the  Helmholtz equation. 
Therefore, for $r>h$ the electrostatic
potential  still has a Yukawa form, but with an undetermined
prefactor, 
\begin{equation}
\label{23}
\rho_q(r)= \frac{A e^{-\kappa r}}{ r}\;.
\end{equation}
The coefficient $A$ can be obtained from 
the condition of continuity of
$\rho_q(r)$ across $r=h$.  We find
\begin{equation}
\label{24}
A= -q \rho h e^{\kappa h}\;.
\end{equation}
The charge neutrality requires that
\begin{equation}
\label{25}
q-\frac{4 \pi }{3} q\rho h^3 +4 \pi \int_h^\infty r^2dr \rho_q(r)=0 \;.
\end{equation}
The integral can be performed explicitly and the resulting equation
solved to determine the size of the correlation hole
\begin{equation}
\label{26}
h=d[\omega(\Gamma)-1]/\sqrt{3 \Gamma} \;,
\end{equation}
where 
\begin{equation}
\label{27}
\omega(\Gamma)=\left[1+(3 \Gamma)^{3/2}\right]^{1/3} \;.
\end{equation}

The size of the correlation hole is a monotonically  increasing function
of the coupling strength.  For high temperatures, small couplings,
$h \approx \lambda_B$.  This is exactly what one might have expected,
since on this length scale the  electrostatic repulsion becomes 
comparable to the thermal energy.  When the temperature is
lowered the kinetic energy diminishes and the particles are strongly
scattered by the electrostatic repulsion, causing increase
in the size of the correlation hole.  For very low temperatures, 
large $\Gamma$, the size of a correlation hole is equal to the
average spacing between the particles, $h \approx d$.  This is, once
again, the correct limiting behavior since at low temperatures the ions 
tend to keep as far away as possible from their neighbors. 
For $r>h$ the electrostatic potential 
follows directly from Eq.~(\ref{21}),
\begin{equation}
\label{28}
\phi_>(r)=\frac{4\pi q \rho h e^{-\kappa (r-h)}}{\epsilon \kappa^2 r} \;.
\end{equation}
For $r \le h$ the Poisson equation with a uniform
background charge $\rho_q=-q \rho$ must be solved.  
The solution is easily found to be
\begin{equation}
\label{29}
\phi_<(r)=\frac{q}{\epsilon r}+\frac{2 \pi q \rho r^2}{3 \epsilon}+\psi \;.
\end{equation}
The induced potential $\psi$ can be determined from the  condition
of continuity of electrostatic potential $\phi_>(h)=\phi_<(h)$, which
reduces to
\begin{equation}
\label{30}
\psi=-\frac{k_B T}{2q}\{[1+(3 \Gamma)^{3/2}]^{2/3}-1\}\;.
\end{equation}
The Debye charging 
process Eq.~(\ref{6}), 
yields the electrostatic free energy per particle
\begin{eqnarray}
\label{31}
\frac{\beta F^{el}}{ N}= \nonumber \\
\frac{1}{4}\left[1-\omega^2+\frac{2 \pi}{3\sqrt{3}}+
\ln\left(\frac{\omega^2+\omega+1}{3}\right)-\frac{2}{\sqrt{3}}
\tan^{-1}\left(\frac{2\omega+1}{\sqrt{3}}\right)\right] \;,
\end{eqnarray}
where $\omega(\Gamma)$ is given by Eq.~(\ref{27}).
In the limit of high temperatures, $\Gamma \rightarrow 0$,
Eq.(\ref{31}) reduces to the Debye limiting law, Eq.~(\ref{7a}). 
Furthermore, the free energy agrees with the Monte Carlo simulations with
an error of less than $10\%$ over a wide range of coupling strengths,
$0<\Gamma<80$. 
This suggests that the inclusion of a correlation
hole into the Debye-H\"uckel theory captures most of the essential physics
of the one component plasma.

Some caution must be taken when using the $OCP$ to model real physical systems.
For $\Gamma>3$,  the isothermal compressibility  and pressure 
of the $OCP$ become negative~\cite{BaHa80}. This
is a consequence of treating the background as a rigid entity and neglecting
its pressure.  How this can be corrected, in practice, depends on the kind
of problem that one wants to study.  If one wants to use the $OCP$ to
model dense ionized matter, the suitable background is the degenerate electron
gas.  When the free energy of background is added to the $OCP$ the
pressure and the compressibility remain non-negative for all 
values of $\Gamma$.  An
alternative approach was suggested by Weeks, who defined the $OCP$ as the
classical ``dense-point limit'' of a two component plasma~\cite{We81}.  
In this limit,
the background is treated as an infinitely dense cloud of point 
particles each 
carrying an infinitesimal charge $e$, so that $e \rho_{back}$ 
remains constant.
Presence of such background does not affect the electrostatics
of the $OCP$, but regularizes its pressure and isothermal compressibility,
making them non-negative.
An interesting byproduct of this analysis is the conclusion that the
freezing  of the $OCP$ occurs without any  change in density, i.e. 
the volume per particle in the fluid and the solid phases is the 
same~\cite{We81}.  
The freezing transition
happens at $\Gamma \approx 180$ and the resulting 
solid phase has the $BCC$ structure~\cite{PaHa73,BaHa80,RoYoWi83}.          

\subsection{Confined one component plasma}
\label{Cocp}

In 1971 Crandall and Williams suggested that electrons 
trapped on the surface of liquid helium $^4$He can crystallize,
forming a two dimensional Wigner crystal~\cite{CrWi71}. Eight years later this
order-disorder transition was observed experimentally 
by Grimes and Adams~\cite{GrAd79}. 
In this system
electrons obey the classical mechanics, since the Fermi energy is much 
smaller than $k_BT$.  Similar crystallization can
occur  in the inversion layer near the surface of a semiconductor, 
however, 
in this case the quantum effects are important and
the electrons form a degenerate quantum gas~\cite{Ch71}.

The trapped electrons above the liquid $^4$He can be modeled as a
confined quasi-two-dimensional plasma of particles interacting by
$1/r$ potential.  This model is also  appropriate for
the study of correlations between the condensed counterions
on the surface of colloidal particles.  

The average spacing between the confined electrons is 
$d=(\pi \sigma)^{-1/2}$, where $\sigma$  is the average surface density, 
$\sigma=N/A$. The dimensionless quantity parameterizing the strength
of electrostatic interactions is $\Gamma=q^2/\epsilon k_B T d$.
For an infinitesimally thin layer separating two mediums
of dielectric constants $\epsilon_1$ and $\epsilon_2$, the important parameter
is the average
dielectric constant $\epsilon=(\epsilon_1+\epsilon_2)/2$.
It  has been observed in computer simulations~\cite{GaChCh79} 
that the 2d $OCP$ 
crystallizes into triangular Wigner crystal for $\Gamma > 130$.
This value is also in a close agreement with the experiments of
Grimes and Adams.

We can gain much insight into thermodynamics of 2d
$OCP$ using the, now familiar, Debye-H\"uckel theory. 
Our model consists
of a plasma of point particles of charge $q$ and  of a neutralizing
background,  confined to an interface located at $z=0$ 
between the two dielectric half-spaces.  
For $z<0$ the dielectric constant is $\epsilon_1$ and
for $z>0$ the dielectric constant is $\epsilon_2$. Since the
half-spaces do not contain any free charges, the electrostatic
potential everywhere satisfies the Laplace equation $\nabla^2 \phi=0$.
The electrostatic free energy is obtained by fixing one particle
and calculating the induced potential resulting from  the 
redistributions of other ions in the $z=0$ plane.  It is
convenient to adopt the cylindrical coordinate system, 
$(\varrho, \varphi, z)$, so that the fixed ion is located at
$\varrho=0, \;z=0$. Using the 
azimuthal symmetry and 
the fact that the electrostatic potential  vanishes  
at infinity,  
the solution to Laplace equation can be written as~\cite{FlBaLe01}
\begin{equation}
\phi_{1}(\varrho,z) = \int_{0}^{\infty} A_{1}(k) 
J_{0}(k\varrho){\bf{\mbox{\large{\em e}}}}^{k z}\; dk\;\;\;\; 
for \;\;\; z<0 \;,
\label{32}
\end{equation}
\noindent and
\begin{equation}
\phi_{2}(\varrho,z) = \int_{0}^{\infty} A_{2}(k) 
J_{0}(k\varrho){\bf{\mbox{\large{\em e}}}}^{-k z}\; dk\;\;\;\; 
for \;\;\; z>0 \;\;\;,
\label{33}
\end{equation}
\noindent where $J_{0}(x)$ is the Bessel function of order zero.

The functions $A_1(k)$ and $A_2(k)$ can be determined 
from the boundary conditions, which are:
continuity of the electrostatic  potential, 
\begin{equation}
\phi_{2}(\varrho,0)=\phi_{1}(\varrho,0) \;\;,
\label{34}
\end{equation}
\noindent and discontinuity of the
displacement field across the $z=0$ plane. The discontinuity  
results from  the 
inhomogeneous distribution of interfacial charge induced by the fixed ion, 
\begin{equation}
\left[ \epsilon_2{\bf E}_{2}(\varrho, z) - \epsilon_1{\bf E}_{1}(\varrho,z)  
\right] \cdot \hat n = 4\pi \sigma_{q}(\varrho) \;\;.
\label{35}
\end{equation}
From charge neutrality the {\it average} interfacial charge
is zero  so that  $\sigma_{q}(\varrho)$ is the 
result of ionic correlations,
\begin{equation}
\sigma_{q}(\varrho)=\frac{q\delta(\varrho)}{2\pi \varrho} -q\sigma +
q\sigma e^{-\beta q \phi(\varrho,0)}\;. 
\label{36}
\end{equation}
The first term of Eq.~(\ref{36}) is the surface charge density of
the fixed ion, the second term is due to the  uniform negative 
background, while the last term is the surface charge density 
of ions confined to the interface. 
We have, once
again, approximated the potential of mean force by the electrostatic
potential.  In the spirit of Debye-H\"uckel theory we shall now
linearize the Boltzmann factor.  The surface charge density 
becomes 
\begin{equation}
\sigma_{q}(\varrho)=\frac{q\delta(\varrho)}{2\pi \varrho}
-\frac{\epsilon \phi(\varrho,0)}{2\pi\lambda_{GC}} \;, 
\label{37}
\end{equation}
where 
\begin{equation}
\lambda_{GC}=\frac{k_B T \epsilon}{2\pi q^2 \sigma} 
\label{37a}
\end{equation}
is the Gouy-Chapman length.

The continuity of electrostatic potential requires that 
$A_1(k)=A_2(k)$. Substituting Eqs.(\ref{32}) and (\ref{33}) into
Eq.(\ref{35}) and using Eq.(\ref{37}), we find the electrostatic
potential over the full range $-\infty<z<\infty$ to be
\begin{equation}
\phi (\varrho,z)=\frac{q}{\epsilon}\int_{0}^
{\infty}\frac{k}{k+\lambda_{GC}^{-1}}J_{0}(k\varrho)
{\bf{\mbox{\large{\em e}}}}^{-k\vert z \vert} dk \;.
\label{38}
\end{equation}
For $z=0$ the integral can be performed explicitly yielding
\begin{equation}
\label{39}
\phi( \varrho,0 )= \frac{q \tau_{0}(\varrho/\lambda_{GC}) }
{ \epsilon \varrho } \;, 
\end{equation}
where the functions $\tau_{\nu}(x)$ are defined as~\cite{VeBl97},
\begin{equation}
\tau_{\nu}(x) = 1 - \frac {\pi x^{1-\nu}}{2} [H_{\nu}(x)-N_{\nu}(x)] \;, 
\label{40}
\end{equation}
with $H_{\nu}(x)$ and $N_{\nu}(x)$ being the 
Struve and the Bessel functions 
of order $\nu$, respectively.  
For large values of $x$, $\tau_0\approx 1/x^2$, so that
asymptotically, 
\begin{equation}
\phi(\varrho,0) \approx \frac{q \lambda_{GC}^2}{\epsilon\varrho^3} \;. 
\label{41}
\end{equation}
We conclude that in the case of a confined plasma 
there is no exponential screening,
instead the electrostatic potential is purely algebraic and 
has the form of 
a dipole-dipole  interaction~\cite{BaHa80}.

There is a well known argument in the condensed matter physics 
going back all the way to Bloch~\cite{Bl30}, 
Peierls~\cite{Pe35} and Landau~\cite{La37} in the  1930's, 
which states that a continuous
symmetry can not be broken in two dimensions.  This means that
there can not exist a true two dimensional crystalline order, 
since it requires breaking translational
symmetry.  The argument was made rigorous by Mermin, who proved
it for particles interacting by  short-ranged potentials~\cite{Me68}.
It is quite simple to see how this conclusion
arises. Suppose that there is a 2d crystal, 
one can then calculate the mean-square 
displacement $\delta^2$ of one particle from its
equilibrium  position due to thermal fluctuations.
It is found that $\delta^2 \sim T \ln L$, where $L$ is the characteristic
crystal size. For $L\rightarrow \infty$, the mean square displacement 
diverges for any finite temperature,  
implying that in thermodynamic limit a 2d crystal
is unstable to thermal fluctuations. 
Although there is no true long-range order in two dimensions for systems
with short-range forces, there exists a pseudo-long-range order
characterized by an algebraically decaying correlation functions.
It is not clear, however, to what extent this conclusion applies to 
the 2d $OCP$, 
whose particles interact by a long-ranged $1/r$ potential. 
Certainly in this case Mermin's proof is no longer valid. 
However, since the effective
interaction potential inside a 2d $OCP$  decays as 
$1/r^3$, which is short-ranged in two dimensions, 
suggests that there should not be
any long-range order.  
Whether there is a true long-range order or a 
pseudo-long-range order for a 2d
$OCP$ remains uncertain.  Simulations  find that 
for $\Gamma\approx 130$ there is a crystallization
transition.  It is, however, difficult to say whether
the crystalline state has a true long-range order or a pseudo-long-range 
order~\cite{GaChCh79}. It is also unclear if the transition is of first
order or continuous, belonging to the Kosterlitz-Thouless universality
class~\cite{KoTh73,Ne83}. Existence of the thermodynamic
limit for confined 2d plasmas can also be attributed to the effective 
renormalization
of the interaction potential from a non-integrable $1/r$ to integrable
(in two dimensions) $1/r^3$ form.

The Helmholtz free energy of a 2d plasma can be obtained directly from 
Eq.~(\ref{39}). We need to know the induced potential felt
by the central ion due to other particles.   
In the limit $\varrho\rightarrow 0$, the  electrostatic potential
reduces to
\begin{equation}
\label{42}
\phi( \varrho, 0) \approx  \frac{q}{ \epsilon \varrho } + 
\frac{q}{\epsilon \lambda_{GC}} \ln(\varrho/2\lambda_{GC})\;. 
\end{equation}
The first term of this expression is the potential produced by
the central ion, while the second term is the induced potential
felt by the fixed ion.  We note
that the induced potential is actually divergent in the limit 
$\varrho\rightarrow 0$.  This is the consequence of the 
failure of linearization of the Poisson-Boltzmann equation. 
This deficiency can be corrected in the same way as was
done for the 3d $OCP$, by introducing a correlation hole of radius $h$
around each particle.
Unfortunately in the present geometry
this leads to calculations which are no longer tractable analytically.
From our study of the 3d $OCP$ we can, however, make some reasonable
approximations. In the limit of high temperatures, small $\Gamma$,
the size of the correlation hole should  be such that the electrostatic and the
thermal energies become approximately equal,
\begin{equation}
\label{43}
\frac{q^2}{\epsilon h} \approx k_BT \;.
\end{equation}
This means that  $h\approx\lambda_B$.  We can use this value as the
short-distance cutoff in the calculation of free energy. The induced
potential then becomes
\begin{equation}
\label{44}
\psi \approx \frac{q}{\epsilon \lambda_{GC}} \ln(\lambda_B/2\lambda_{GC}) \;.
\end{equation}
The free energy is obtained through the usual Debye charging 
process, Eq.~(\ref{6}).  Recalling 
that $\lambda_B(\lambda q)=\lambda^2\lambda_B(q)$ and  
$\lambda_{GC}(\lambda q)=\lambda_{GC}(q)/\lambda^2$, where 
$\lambda$ is the charging parameter,  in the
limit of high temperatures $\Gamma \rightarrow 0$, the reduced free
energy per particle is found to be
\begin{equation}
\label{45}
\frac{\beta F^{el}}{N} \approx \Gamma^2 \ln(\Gamma) \;.
\end{equation}
Eq.~(\ref{45}) is precisely the leading order term of the resumed
virial expansion obtained by Totsuji~\cite{To75,To76}.

For low temperatures, the $OCP$ crystallizes into a triangular
lattice. The Madelung energy of this lattice is,
\begin{equation}
\label{47}
\frac{\beta U}{N}=-1.106103 \Gamma  \;.
\end{equation}
This equation provides a surprisingly good fit not only for the
free energy of solid, but also for the free
energy of fluid
at sufficiently high values of $\Gamma$.  Comparing to the
results of the Monte Carlo simulations~\cite{GaChCh79} we find that for
$\Gamma=5$ the error accrued from using Eq.~(\ref{47}) to calculate
the total electrostatic 
free energy is about $30\%$.  For $\Gamma=20$ this error
drops to $11\%$ and for $\Gamma=50$ it goes down to $6\%$. 
Recalling that the crystallization 
transition occurs at $\Gamma\approx 130$, we see that the Eq.~(\ref{47})
works well into the fluid phase.
It is reasonable, therefore, to approximate the electrostatic free energy
of a fluid for $\Gamma>5$ by
\begin{equation}
\label{48}
\frac{\beta F^{el}}{N}=-1.106103 \Gamma  \;.
\end{equation}
The reason why the electrostatic free energy of a fluid
is so well approximated by the free energy
of the crystal, is a consequence of strong electrostatic 
correlations. 

\section{Asymmetric systems}

Up to now we have considered only symmetric plasmas and electrolytes.
In practice, however, it is unlikely that both cations and
anions have exactly the same size and magnitude of charge.  
It is, therefore, important to explore the thermodynamics of 
a general $Z:1$ electrolyte
in which cations have charge $Zq$ and diameter $a_c$, while anions
have charge $-q$ and diameter $a_a$. Unfortunately, as soon as
the asymmetry is introduced, the internal inconsistency enters into
the Poisson-Boltzmann equation~\cite{On33}.   Recall that 
the cation-anion correlation
function can be expressed in terms of the 
potential of mean force $w_{+-}$   
\begin{equation}
\label{49}
g_{+-}(r)=e^{-\beta w_{+-}(r)} \;.  
\end{equation}
The $w_{+-}(r)$ is the work needed to bring
cation and anion from infinity to separation $r$ inside an electrolyte.  
Clearly
this work is invariant under the permutation of particle
labels $w_{+-}(r)=w_{-+}(r)$.  This means that
\begin{equation}
\label{50}
g_{+-}(r)=g_{-+}(r) \;. 
\end{equation}
The Poisson-Boltzmann equation, which serves as the basis, for the
Debye-Huckel theory, approximates the potential of mean force by 
$w_{+-}(r)=q_-\phi_+(r)$.  The
self consistency condition, Eq.~(\ref{50}), then requires that
\begin{equation}
\label{51} 
q_+\phi_-(r)=q_- \phi_+(r) \;.
\end{equation}
Because of the non-linear nature of the $PB$ equation this
condition can not be satisfied except for symmetric electrolytes.  The 
linearization prescription intrinsic to the Debye-H\"uckel theory allows
Eq.~(\ref{51}) to hold for ions of different valence, but with
the {\it same } ionic diameter, $a_c=a_a$.  

We see that as soon as the symmetry between the cations and anions is 
broken the physics and the mathematics of the problem becomes significantly
more complex.  In the limit of very large asymmetries, $Z \rightarrow \infty$
and $a_c \gg a_a$ a new simplification, however, enters into the game.

\subsection{Colloidal suspensions}

A typical colloidal suspension often studied experimentally consists of
polystyrene sulphonate spheres of diameter $10nm-1\mu m$ and $10^3-10^4$
ionizable surface groups.  Because of the large surface charge, the
colloidal particles tend to repel each other,  forming 
crystals, even at fairly low volume fraction of less than $10\%$.
Using the periodic structure of the lattice, 
the thermodynamics of a colloidal crystal can be 
studied fairly straightforwardly.  Each colloidal particle 
can be thought to be confined to a Wigner-Seitz (WS) polyhedral cell.
A further approximation replaces the polyhedral
$WS$ cell by a sphere~\cite{AlChGr84}.

\subsection{Colloidal lattices}

We shall model the colloidal particles as hard spheres of radius $a$ carrying
$Z$ ionizable groups  of charge $-q$ distributed uniformly 
on the surface.  The counterions
will be idealized as point particles of charge $+q$.  The suspension of
$N_p=\rho_p V$ polyions and $N_c=ZN_p=\rho_cV$ counterions is confined to a
volume $V$.  As usual, the solvent will 
be treated as a uniform continuum of
dielectric constant $\epsilon$.  For sufficiently large
polyion concentrations colloidal suspension crystallizes.
Using the lattice symmetry, we restrict our attention to
{\it one} colloidal particle and its 
counterions inside a spherical $WS$ cell of radius $R$ such that
\begin{equation}
\label{52} 
\rho_p=\frac{1}{\frac{4 \pi}{3} R^3}\;.
\end{equation}
Using the statistical mechanics it is possible to show that
the osmotic pressure inside a cell is proportional to 
the concentration
of counterions at the cell boundary~\cite{WeJoLi82}, 
\begin{equation}
\label{53} 
\beta P=\rho_c(R)\;.
\end{equation}
The thermodynamics of a crystalline  colloidal suspension now
reduces to the calculation of the distribution of counterions
inside a $WS$ cell. This can be done using a simple 
mean-field picture.  The electrostatic potential inside a
$WS$ cell satisfies the Poisson equation (\ref{0}) with the
counterion charge density approximated by the normalized 
spherically symmetric Boltzmann distribution,
\begin{equation}
\label{54} 
\rho_q(r)=Zq N_p \frac{e^{-\beta q \phi(r)}}{4 \pi \int_a^R r^2 dr 
e^{- \beta q \phi(r)} }\;.
\end{equation}

The non-linear Poisson-Boltzmann equation can be solved numerically to yield
the electrostatic potential and the distribution of counterions inside the
cell. In practice it is more convenient to work with  the electric field
\begin{equation}
\label{55} 
\bold{E(r)}=-\nabla \phi(\bold r)\;.
\end{equation}
The Poisson equation can then  be
rewritten as
\begin{equation}
\label{56} 
\nabla \cdot \bold{E(r)} = \frac{4 \pi}{\epsilon }
[\rho_q(\bold r)+ q_p(\bold r)] \;,
\end{equation}
where $q_p(\bold r)$ is the polyion charge density,
\begin{equation}
\label{56a} 
q_p(\bold r)=-\frac{Zq}{4 \pi a^2} \delta(\bold{ |r|}-a)] \;.
\end{equation}
To simplify the calculations we have uniformly smeared the charge of the
polyion over its surface.
Integrating both sides of Eq.~(\ref{56}) and taking advantage of the
divergence theorem, the electric field at distance $r$ from the polyion is
\begin{equation}
\label{57} 
E(r) =- \frac{1 }{\epsilon r^2  }\left[Zq-\alpha(r)\right] \;,
\end{equation}
where
\begin{equation}
\label{58} 
\alpha(r)=\int_{|\bold{r'}|<r} d^3{\bold r'} \rho_q(\bold r') \;
\end{equation}
is the  counterion charge inside a sphere of radius $r$ centered
on the colloidal particle. Using the gauge in which
$\phi(a)=0$ the electrostatic potential is
\begin{equation}
\label{59} 
\phi(r)=-\int_a^r dr E(r)\;
\end{equation}
and the Poisson-Boltzmann equation reduces to an integral equation for 
the electric field. Note that Eq.~(\ref{57}) naturally incorporates the
boundary conditions
\begin{equation}
\label{60} 
E(a)=-\frac{Zq}{\epsilon a^2}\;
\end{equation}
and 
\begin{equation}
\label{61} 
E(R)=0\;.
\end{equation}
Eq.~(\ref{57}) can be solved iteratively to yield the counterion
density profile from which all other thermodynamic functions 
are straightforwardly determined.

For aqueous colloidal lattices with  monovalent counterions, the 
Poisson-Boltzmann equation is in excellent agreement with the 
experiments
and simulation.
The $PB$ equation, however, does not account for the
correlations between the counterions
and breaks down for low dielectric solvents or for aqueous suspensions
with multivalent ions.  Fortunately, in the case of colloidal lattices,
it is fairly straightforward to account for these effects using the
density functional theory.

\subsection{Density functional theory $(DFT)$ for colloidal lattices}
\label{DFT}
We shall now  construct the Helmholtz free energy functional 
for a Wigner-Seitz
cell of a colloidal lattice. The Helmholtz free energy is a 
functional of the
average local counterion concentration
\begin{equation}
\label{61a} 
\rho_c(\bold r)=\langle \sum_{i=1}^{N_c} \delta(\bold r -\bold r_i) \rangle\;,
\end{equation}
where the brackets denote a Boltzmann average over all 
the particle positions.
The free energy consists of electrostatic
and entropic contributions.  The entropic contribution is simply that
of an inhomogeneous ideal gas,
\begin{equation}
\label{61b} 
\beta F^{ent}[\rho_c(\bold r)]=\int d^3\bold r \, \rho_c(\bold r)\{ 
\ln[\rho_c(\bold r) \Lambda^3]-1\}\;.
\end{equation}
The electrostatic contribution is the result of Coulomb interactions
between the counterions and the polyion, as well as the
self-energy of the ionic cloud,
\begin{equation}
\label{61c} 
F^{el}[\rho_c(\bold r)]=q\int d^3\bold r \, d^3\bold {r'}\,
\frac{q_p(\bold {r'})\rho_c(\bold r)}{\epsilon|\bold{r}-\bold{r'}|}
+\frac{q^2}{2}\int d^3\bold r \, d^3\bold {r'}\,
\frac{\rho_c(\bold {r'})\rho_c(\bold r)}{\epsilon|\bold{r}-\bold{r'}|}\;.
\end{equation}
Eq.~(\ref{61c}) is the 
mean-field approximation for the total
electrostatic free energy.  It does not account for the
electrostatic correlations between the 
counterions surrounding the colloidal particle. Clearly, if
there is a counterion present at position $\bold r$, there is 
a reduced probability of finding another counterion in its vicinity.
This information is not included in Eq.~(\ref{61c}). 
One can attempt to account for the electrostatic correlations 
using the local 
density approximation $(LDA)$,
\begin{equation}
\label{61d} 
\beta F^{cor}[\rho_c(\bold r)]=\int d^3\bold r \,\rho_c(\bold r) f_{cor}[\rho_c(\bold r)] \;.
\end{equation}
The correlational
free energy can be approximated by the free energy of a three dimensional
one component plasma Eq.~(\ref{31}),
\begin{eqnarray}
\label{61e} 
f_{cor}[\rho_c(\bold r)] \approx f_{ocp}[\rho_c(\bold r)]= \nonumber \\
\frac{1}{4}\left[1-\omega^2+\frac{2 \pi}{3\sqrt{3}}+
\ln\left(\frac{\omega^2+\omega+1}{3}\right)-\frac{2}{\sqrt{3}}
\tan^{-1}\left(\frac{2\omega+1}{\sqrt{3}}\right)\right] \;,
\end{eqnarray}
with 
\begin{equation}
\label{61f} 
\omega[\rho_c(\bold r)]=
\left[1+(3 \Gamma[\rho_c(\bold r)])^{3/2}\right]^{1/3}\;,
\end{equation}
where the local coupling strength is,
\begin{equation}
\label{61g} 
\Gamma[\rho_c(\bold r)]=\lambda_B[4 \pi \rho_c(\bold r)/3]^{1/3}\;.
\end{equation}
Although Eq.~(\ref{61e}) was derived for a plasma in a uniform
neutralizing background, it can also be used to approximate
the correlational  electrostatic
free energy of counterions confined in a $WS$ cell.  The role of a 
neutralizing background being played by the confining electric field
produced by the colloidal particle.

The equilibrium charge distribution is determined from  
the minimization of the total Helmholtz free energy
\begin{equation}
\label{61h} 
F=F^{ent}+F^{el}+F^{cor}\;
\end{equation}
subject to  constraint of particle conservation
\begin{equation}
\label{61i} 
\int d^3\bold r \,\rho_c(\bold r)=N_c\;.
\end{equation}
This is equivalent to minimization of the grand potential
\begin{equation}
\label{61j} 
\Omega=F-\mu_c N_c\;,
\end{equation}
where  the chemical potential of counterions $\mu_c$ is the 
Lagrange multiplier.  Performing the calculation, leads to
the equilibrium counterion density profile
\begin{equation}
\label{61k} 
\rho_c(r)=N_c \frac{e^{-\beta q \phi(r) - 
\beta \mu^{ex}(r)}}{4 \pi \int_a^R r^2 dr 
e^{- \beta q \phi(r)-\beta \mu^{ex}(r)} }\;,
\end{equation}
where the excess chemical potential is,
\begin{equation}
\label{61l} 
\mu^{ex}(\bold r)=\frac{\delta F^{cor}}{\delta \rho_c(\bold r)}\;.
\end{equation}
We see that in the absence of correlations the density profile
reduces to the Boltzmann distribution Eq.(\ref{54}). 
The counterion density, which enters into the expression for
the excess chemical potential, 
can be expressed in terms of the electric field,
\begin{equation}
\label{61m} 
\rho_c(\bold r)=\frac{\epsilon}{4 \pi q} \nabla \cdot \bold E\;,
\end{equation}
which due to spherical symmetry simplifies to
\begin{equation}
\label{61n} 
\rho_c(r)=\frac{\epsilon}{4 \pi q r^2} \frac{\partial (r^2E)}{ \partial r}\;.
\end{equation}
Substituting Eq.(\ref{61k}) into Poisson equation (\ref{56}) and using  
Eqs.~(\ref{59}) and (\ref{61n}) leads to an
integro-differential equation for the 
electric field, Eq.(\ref{57}).

Unfortunately,
for large colloidal charges this equation has no 
stable solutions~\cite{PeNoJo90,StRo90,Gr91}. There
is an unbounded increase in the concentration of
counterions in the vicinity of the colloidal surface 
resulting from the failure
of the local density approximation~\cite{BaDeHo00}.
To overcome this difficulty Groot and ~\cite{Gr91} has suggested use of a 
weighted density approximation $(WDA)$, which has proven quite
successful in its application to other problems of the condensed 
matter physics~\cite{Ta85,CuAs85}.
Within this approach the correlational free energy is given
by  
\begin{equation}
\label{61p} 
F^{cor}[\rho_c(\bold r)]=\int d^3\bold r \,\rho_c(\bold r) 
f_{cor}[\bar \rho_c(\bold r)] \;,
\end{equation}
where $\bar \rho_c(\bold r)$ is the {\it average} local density,
\begin{equation}
\label{61q} 
\bar \rho_c(\bold r)=\int d^3\bold r' 
w(|\bold r-\bold r'|) \rho_c(\bold r)  \;.
\end{equation}
The weight function $w(r)$ can be determined from
the thermodynamic requirement that
\begin{equation}
\label{61r} 
\frac{\delta^2 \beta F}{\delta \rho_c(\bold r)\delta \rho_c(\bold r') }=
\frac{\delta(\bold r-\bold r')}{\rho_c(\bold r)}-C_2(|\bold r-\bold r'|) \;,
\end{equation}
where $C_2(|\bold r-\bold r'|)$ is the direct correlation function.
In particular this equation must hold in the limit of a homogeneous
$OCP$, the direct correlation function for which
can be obtained  using the theory developed
in Section \ref{OCP}~\cite{Gr91},
\begin{equation}
\label{61s} 
 C_2(r)=-\frac{\lambda_B}{h} \;\;\; if \;\;\; r \le h \;,
\end{equation}
and 
\begin{equation}
\label{61t} 
 C_2(r)=-\frac{\lambda_B}{r} \;\;\; if \;\;\; r>h \;.
\end{equation}
Performing the calculation,
we find that the weight function is well approximated by~\cite{Gr91}
\begin{equation}
\label{61u} 
w(r)=\frac{3}{2 \pi h^2}\left(\frac{1}{r}-\frac{1}{h}\right)\Theta(h-r) \;,
\end{equation}
where $\Theta(x)$ is the Heaviside step function, $\Theta(x)=1$ for $x>0$
and $\Theta(x)=0$ for $x<0$. The $WDA$ is significantly more 
computationally demanding than the $LDA$.  
Its advantage, however, is the numerical 
stability for all values of the colloidal charge. The results 
based on the $WDA$ 
are in excellent agreement
with the Monte Carlo simulations.  The $WDA$ gives us 
a good handle
on the thermodynamics of colloidal lattices. 
At lower concentrations, when the crystalline structure
has melted, the situation unfortunately is no longer so clear cut.  
In this case a
simple picture based on the Wigner-Seitz cell is  not sufficient
and new methods must be developed~\cite{LeBaTa98,TaLeBa98}.  
Unfortunately the standard techniques
of the liquid state theory based on  integral equations 
are powerless in the case of highly
asymmetric colloidal systems. The field theoretic methods 
also fail
when applied to
this difficult problem, Fig. \ref{Fig0}. 
Furthermore, even the experimental situation is far from clear. 
Ise and coworkers
claim to have seen stable clusters of colloidal particles in
highly deionized colloidal suspensions.  Tata {\it et al.} even report
an observation of a full equilibrium 
vapor-liquid-like phase separation~\cite{TaRaAr92}.  
These experiments, however,
have been challenged by Palberg and W\"urth, who demonstrated that the
phase separation 
observed by  Tata {\it et al.} was the result of non-equilibrium
salt gradients produced by the ion exchange resin~\cite{PaWu94,TaAr94}. In the
colloidal science community the possibility of a liquid-vapor phase
separation in highly deionized colloidal suspensions has met
with a large amount of scepticism.  The usual argument against
the phase transition is based on the Derjaguin-Landau-Verwey-Overbeek 
$(DLVO)$ colloidal pair  potential~\cite{DeLa41,VeOv48}.

It is easy to understand
the nature of the $DLVO$ potential based on the Debye-H\"uckel theory.
If the size of colloidal particles is shrunk to
zero, $a\rightarrow 0$, then due to  screening by counterions,                 
the interaction energy between two ``point''
colloids would be of a Yukawa form, 
\begin{equation}
\label{62} 
V_{0}(r)=(Zq)^2 \frac{e^{-\kappa r}}{\epsilon r}\;, 
\end{equation}
where the inverse Debye length is 
\begin{equation}
\label{63} 
\xi_D^{-1}\equiv \kappa=\sqrt{\frac{4 \pi Z q^2 \rho_p}{k_B T \epsilon}}\;. 
\end{equation}
Now, consider the electrostatic potential outside
the fixed colloidal particle of radius $a$ and charge $-Zq$,  Eq.~(\ref{5})
\begin{eqnarray}
\label{64}
\phi_>(r)=-\frac{Zq \theta(\kappa a) e^{-\kappa r}}{\epsilon r} \;, 
\;\;\;\; 
\theta(x)=\frac{e^{x}}{(1+x)}\;.
\end{eqnarray}
Evidently the factor $\theta(\kappa a)$ accounts for the fact that
screening starts only outside the cavity, $r>a$. We also
can think of Eq.~(\ref{64}) as the potential  
of a point particle with an effective
charge $Q_{p}=Zq \theta(\kappa a)$.  An advantage
of this alternative point of view is that the interaction
energy for two ``point'' particles 
is simply given by Eq.~(\ref{62}) 
with $Zq \rightarrow Q_{p}$. This leads directly to 
the famous  $DLVO$ potential
\begin{equation}
\label{65} 
V_{DLVO}(r)=(Zq)^2\theta^2(\kappa a)\frac{e^{ -\kappa r}}{\epsilon r}\;. 
\end{equation}
This potential is purely repulsive~\cite{RoKrGr88}, 
which naively suggests that a charged 
colloidal suspension is stable against a
liquid-gas phase separation.
Sogami and Ise, therefore, have argued that the $DLVO$
potential  must be incorrect,
since it cannot account for the 
inhomogeneities observed experimentally~\cite{SoIs84}.  
In its stead, they proposed a different interaction 
potential derived on the basis of
the Gibbs free energy. The potential found by Sogami and Ise contains
a minimum~\cite{SoIs84}, which implies that at short enough separations the
two like-charged colloidal particles attract!
What is most surprising is that the
attraction appears even for monovalent counterions, i.e
in the absence of strong correlations between the colloidal double
layers.
Furthermore,  water is an incompressible fluid so that it is
difficult to see how a change of paradigm from  Helmholtz to the 
Gibbs free energy
can lead to such a profound modification of  the interaction potential.
Inconsistency in the results based on the Helmholtz and the Gibbs 
free energies has been carefully 
reexamined by Overbeek, who
has traced the discrepancy to a flaw in the Sogami and Ise's 
calculations~\cite{Ov87}.

It is important to stress that the repulsive two-body 
interactions do not, in general, preclude the possibility of a
liquid-gas phase separation
in a multicomponent fluid.
In fact van Roij and Hansen found,
within  the linearized density functional theory, that it is  
possible  for a colloidal
suspension with polyions interacting by the repulsive $DLVO$ potential
to phase separate into  coexisting liquid and gas phases~\cite{RoHa97}. 
Before entering into the discussion of colloidal fluids it 
is, however, important to introduce a new 
fundamental concept --- the colloidal charge renormalization.

\subsection{Charge Renormalization}

Although the non-linear Poisson-Boltzmann equation can not be
solved analytically for a spherical geometry, 
the numerical solution indicates that the
electrostatic potential far from 
colloidal particle saturates as a 
as a function of the bare colloidal charge~\cite{AlChGr84}. 
This suggests that the thermodynamics
of a highly charged colloidal systems can be based on a linearized
$PB$ equation but with the bare colloidal charge replaced 
by an effective renormalized charge.  
The original concept of colloidal
charge renormalization is due to Alexander et al., but is well 
predated
in the polyelectrolyte literature, where the phenomenon is  known as 
the Manning counterion condensation~\cite{Ma69,Oo71,Ma78}.  

To understand better colloidal charge renormalization, 
let us first consider a
uniformly charged plane at fixed potential $\psi_s$ inside a
salt solution of concentration $c$. The electrostatic potential
at distance $x$ from the plane satisfies the PB equation,
\begin{equation}
\label{65aa}
\frac{d^2\phi(x)}{dx^2}=\frac{8 \pi c q}{\epsilon}\sinh(\beta q \phi) \;.
\end{equation}
Since at the moment we are considering aqueous suspensions
containing only monovalent ions, the electrostatic 
correlations are
insignificant and the mean-field Poisson-Boltzmann approximation
is  sufficient. Multiplying both sides of
Eq.~(\ref{65aa}) by $d\phi/dx$ allows us to perform the first integration.
Since the potential vanishes in the limit $x \rightarrow \infty$, 
we find
\begin{equation}
\label{65ab}
\frac{1}{2}[\phi'(x)]^2=\frac{8 \pi c}{\epsilon \beta}
\left[\cosh(\beta q \phi)-1\right] \;.
\end{equation}
The second integration yields~\cite{RuSaSc89}
\begin{equation}
\label{65ac}
\phi(x)=\frac{2 k_B T}{q} \ln \frac{1+e^{-\kappa x} \tanh(\beta q\psi_s/4)}
{1-e^{-\kappa x} \tanh(\beta q\psi_s/4)} \;,
\end{equation}
where the inverse Debye length is,
\begin{equation}
\label{65aca}
\kappa = \sqrt{8 \pi c \lambda_B} \;.
\end{equation}
In the limit of large surface potentials 
this expression simplifies to 
\begin{equation}
\label{65ad}
\phi(x)=\frac{2 k_B T}{q} \ln \frac{1+e^{-\kappa x}}
{1-e^{-\kappa x}} \;.
\end{equation}
For separations from the plane larger than the Debye length, Eq.~(\ref{65ad})
becomes
\begin{equation}
\label{65ae}
\phi(x)=\frac{4 k_B T}{q} e^{-\kappa x} \;.
\end{equation}
An important observation is that for large surface potentials, 
$\beta q\psi_s/4 \gg 1$, the
electrostatics away from the plane is completely
insensitive to the surface charge density.  

Now, let us consider a highly charged colloidal particle 
of valence $Z$ and radius $a$
inside a symmetric $1:1$ electrolyte of concentration $c$. 
The electrostatic potential at distance $r$ from the center of a colloidal
particle satisfies the $PB$ equation (\ref{2}).  For distances $r>a+\xi_D$
the electrostatic potential is small and the $PB$ equation can
be safely linearized leading to the Helmholtz equation (\ref{3}).  This
can be easily integrated yielding the electrostatic potential,
\begin{equation}
\label{65af}
\phi(r)=A \frac{e^{-\kappa r}}{r} \;,
\end{equation}
where $A$ is the  integration constant.  To find its value, lets restrict
our attention to suspensions in which the $\xi_D \ll a$.  In practice this is
not a very strong restriction.  For salt solutions at physiological
concentrations $\xi_D \approx 8$ \AA $\,$ while the characteristic
colloidal size is on the order of $1000$ \AA.  Even for solutions with
very low salt content, in the $mM$ range, the Debye length is on the
order of $100$ \AA. Under these conditions all the curvature
effects associated with the spherical geometry of colloidal particle
are effectively screened at separations $a+\xi_D < r < 2a$, and
the electrostatic potential is well  approximated by that of a uniformly
charged plane, Eq.~(\ref{65ae}).  Comparing Eqs.~(\ref{65ae}) and 
Eq.~(\ref{65af}) the value of the integration constant follows directly, and
the electrostatic potential at distance $r > a+\xi_D $ 
from the center of colloidal
particle is
\begin{equation}
\label{65ag}
\phi(r)=\frac{4 k_B T a e^{-\kappa (r-a)}}{q r} \;.
\end{equation}
This is the asymptotic solution of the full non-linear $PB$ equation
for $\kappa a \gg 1$.  
Comparing this to the solution of linearized $PB$, Eq.~(\ref{64}), 
it is evident that
the two are identical as long as  
{\it the bare colloidal charge is replaced by the renormalized charge}.  
For
highly charged particles, Eq.~(\ref{65ag}) shows that
the renormalized charge saturates at~\cite{TrBoAu02} 
\begin{equation}
\label{65ah}
Z^{sat}_{ren}=\frac{4 a  (1+\kappa a)}{\lambda_B}\;.
\end{equation}

While the previous analysis was carried out for one colloidal 
particle inside
an electrolyte solution, the concept of 
charge renormalization is quite general
and can be applied to colloidal suspensions 
under various conditions~\cite{QuCaHi00,QuCaHi00,FeFeNi00,FeFeNi01}.
The difficulty of defining the effective charge for suspensions at 
non-zero concentrations resides in the complexity of accounting for the
consequences of colloidal interactions.  The standard practice is to
to study one colloidal particle inside a spherical Wigner-Seitz 
cell whose radius  is determined by the volume fraction of 
colloids~\cite{AlChGr84}.  While this
procedure is fully justified for colloidal lattices, its foundation
is less certain for  fluidized suspensions.  
To find the renormalized charge one numerically solves the full 
non-linear $PB$ equation 
and matches the electrostatic potential to the solution
of the linearized
equation at the cell boundary.  Alternatively, the osmotic pressures
inside the $WS$ cell 
calculated using the non-linear and linear equations are matched
in order to define the effective charge.  One should remember, however,
that while at the level of non-linear $PB$ equation the osmotic pressure is
directly proportional to the concentration of ions at the cell boundary,
Eq.~(\ref{53}), this is not the case for the linearized $PB$ equation.
The various procedures lead to similar values of the renormalized
charge.  In the case of salt-free suspensions, the effective charge
is found to saturate at~\cite{AlChGr84}
\begin{equation}
\label{65ai}
Z^{sat}_{ren} \approx \frac{\chi a}{\lambda_B}\;,
\end{equation}
where  $\chi$ is an approximately linearly increasing function 
of colloidal concentration for suspensions with volume fraction larger than
$1\%$. For suspensions  
with colloidal volume fraction between $1\%$ and $10\%$
the value of $\chi$ varies from around 
$9$ to $15$~\cite{StFaRo96,Be98}.  

\subsection{Colloidal Fluid}
\label{Cfluid}
In this section we will apply the insights gained from the study of
one and two component plasmas to the exploration of
stability of charged colloidal suspensions against a gas-liquid
phase separations. We note that the large size asymmetry between
colloids and counterions leads to very different equilibration
time scales.  On the time scale of polyion motion, the counterions 
are always equilibrated. This suggests that the calculation
of free energy should be done in two stages~\cite{LoHaMa93}.  
First, we shall trace out
the counterion degrees of freedom, leading to effective
many-body interactions between the colloidal particles. Then we will 
use these 
effective interactions to calculate  the colloid-colloid
contribution to the total free energy.  The procedure is similar to
the one used in McMillan-Mayer theory of solutions~\cite{McMa45}.

We shall first calculate the contribution to the total free energy
arising from the polyion-counterion 
interactions~\cite{LeBaTa98,TaLeBa98}.
Consider a suspension in thermal equilibrium. 
While the colloidal particles are more or less  
uniformly distributed throughout the 
solution, the positions of counterions are strongly
correlated with the positions of polyions.  As a leading
order approximation we can, therefore, take the polyion-polyion
correlation function to be 
\begin{equation}
\label{65a} 
g_{pp}=1\; 
\end{equation}
and the polyion-counterion correlation function to be
\begin{equation}
\label{65b} 
g_{pc}=e^{-\beta q \phi(r)}\;. 
\end{equation}
Choosing the coordinate system in such a
way that it is centered on top of one of the colloidal particles,
the electrostatic potential at distance
$r<a$ satisfies the Laplace equation, 
while for distances $r>a$ it
satisfies the Poisson equation Eq.~(\ref{0}).  
Based on Eqs.~(\ref{65a}) and (\ref{65b}) the charge density in the region
$r>a$ can  be approximated by,
\begin{equation}
\label{66} 
\rho_q(r)=-Zq\rho_p+q\rho_c e^{-\beta q \phi(r)}\;. 
\end{equation}
In the spirit of the Debye-H\"uckel theory we shall linearize
the exponential~\cite{DeHu23,LeFi96}.  The distribution
of charge around the colloid reduces to
\begin{equation}
\label{69} 
\rho_q(r)=- \beta q^2 \rho_c \phi(r)\;. 
\end{equation}
For $r>a$ the electrostatic potential,  
therefore, satisfies the Helmholtz equation (\ref{3})
with $\kappa$ given by Eq.~(\ref{63}).  
The solution to this equation is
\begin{equation}
\label{72}
\phi_>(r)=-\frac{Zq \theta(\kappa a )e^{-\kappa r}}{\epsilon r}\;,
\end{equation}
while the solution to the Laplace equation for
$r\le a$ is
\begin{equation}
\label{71}
\phi_<(r)=-\frac{Zq}{\epsilon a (1+\kappa a)}\;.
\end{equation}
The electrostatic energy due to polyion-counterion
interaction is 
\begin{equation}
\label{73}
u_p=\frac{1}{2} \int d^3 \bold r [\rho_q(\bold r)+q_p(\bold r)]\phi(r),
\end{equation}
where $\rho_q(\bold r)$ is the charge density of counterions given
by Eq.~(\ref{69}), and $q_p(\bold r)$  is the 
charge density of a polyion,
\begin{equation}
\label{74}
q_p(\bold r)=-\frac{Zq}{4 \pi a^2} \delta(\bold{ |r|}-a) \;.
\end{equation}
Performing the integration we find
\begin{equation}
\label{75}
u_p=\frac{Z^2q^2}{2 \epsilon  (1+\kappa a)}\left[ \frac{1}{a}-
\frac{\kappa}{2(1+\kappa a) }\right]\;.
\end{equation}
The electrostatic {\it free energy} of a polyion inside the
suspension is obtained using the Debye charging process~\cite{Ma55},
\begin{equation}
\label{76}
{\cal F}_p=\int_0^1 d\lambda \frac{2 u_p(\lambda q)}{\lambda}=
\frac{Z^2 q^2}{2 \epsilon a (1+\kappa a)}\;.
\end{equation}
Note that this free energy is the sum of the polyion self energy 
\begin{equation}
\label{76a}
{\cal F}_p^{self}=\frac{Z^2 q^2}{2 \epsilon a}\;,
\end{equation}
and the solvation energy
\begin{equation}
\label{77}
{\cal F}_p^{solv}=-\frac{Z^2 q^2 \kappa a}{2 \epsilon a (1+\kappa a)}\;,
\end{equation}
which the polyion gains from
being inside the ``ionic sea''. The electrostatic free energy 
due to interaction between all the polyions and counterions is
\begin{equation}
\label{77a}
F^{pc}=-\frac{Z^2 q^2 N_p \kappa a}{2 \epsilon a (1+\kappa a)}\;.
\end{equation}

We have effectively integrated out the counterion degrees of freedom.  This, 
however, leaves us with the effective many-body potentials of interaction
between the colloidal particles. For dilute suspensions, 
the pairwise interaction potential should be the dominant one. 
The two-body interaction potential
can be obtained from the solution of Helmholtz equation for two
colloidal particles~\cite{LiLeFi94,FiLeLi94}.  
At large separations this leads directly to the 
$DLVO$ interaction potential, Eq.~(\ref{65}).
This potential has been extensively tested experimentally and found
to work very well for bulk colloidal suspensions~\cite{CrGr94}. 
Since the $DLVO$ potential
is short ranged, the contribution to the total free energy arising
from the colloid-colloid interaction can be calculated in the spirit
of the 
traditional van der Waals theory, through the second virial term.  A more
sophisticated calculation of the colloid-colloid free energy 
relies on the Gibbs-Bogoliubov variational
bound, 
\begin{equation}
\label{79} 
F^{pp} \le F_0+\langle V_{DLVO}\rangle_0\;, 
\end{equation}
where the reference system is taken to be the fluid of hard spheres, whose 
diameter
plays the role of a variational parameter. The free  energy 
resulting  from the polyion-polyion interaction,
$F^{pp}$,  can be approximated by the lowest variational bound 
of Eq.~(\ref{79}).  
The calculation is somewhat involved, so we refer the interested reader
to the original papers~\cite{MaCa69,FiAs77,RoHa97,Wa00}.

The entropic mixing free energy of colloids and their
counterions is simply that of an ideal gas, 
\begin{equation}
\label{83} 
\beta F^{ent}=
ZN_p[ \ln(Z \rho_p \Lambda^3_c) -1]+ N_p [\ln( \rho_p\Lambda^3_p) -1]\;, 
\end{equation}
where $\Lambda_c$ and $\Lambda_p$ are the de Broglie thermal wavelengths
of counterions and polyions, respectively.

The total free energy of colloidal suspension is 
the free energy needed to solvate colloids
in the sea of other polyions and counterions $F^{pc}+F^{pp}$,
and the free energy of mixing $F^{ent}$,
\begin{equation}
\label{87} 
F=F^{pc}+F^{pp}+F^{ent} \;. 
\end{equation}
The osmotic pressure is
\begin{equation}
\label{87a} 
P=-\frac{\partial F}{\partial V}\Big\arrowvert_{N_p}. 
\end{equation}
It is found that for suspensions with
\begin{equation}
\label{87b} 
{\cal C} \equiv \frac{Z \lambda_B}{a} > 15.2 
\end{equation}
the pressure is not  a convex function of the colloidal
concentration, implying existence of a thermodynamic instability.
At criticality the colloidal volume fraction is around $1\%$. 
The crucial question is whether
this result is reliable? In order to calculate the electrostatic
free energies, we were forced to linearize the Boltzmann factor.
While this is a reasonable approximation at large separations away
from the polyions, 
linearization is clearly invalid in the vicinity
of colloidal surface.  There,  the strong electrostatic interactions
result in an accumulation of counterions 
and the effective polyion charge renormalization. 
Therefore, the linear theory can be used {\it only if} the 
bare colloidal charge is replaced by the effective
renormalized charge,  
$Z \rightarrow Z_{eff}$,  
in all the expressions.  It was  found,
however, that the bare charge does not increase without limit but
saturates at the value given by the Eq.~(\ref{65ai}).  Substituting 
$Z \rightarrow Z_{eff}$, into the definition of ${\cal C}$ Eq.~(\ref{87b}),
we see that ${\cal C}<15$ for all the values of the 
bare charge $Z$ in the critical region. 
The critical threshold, therefore, can not be reached, meaning that 
a deionized {\it aqueous} suspensions with {\it monovalent} counterions 
is stable against a liquid-gas phase
separation  for all colloidal charges and 
sizes.  This conclusion has also been confirmed by more
detailed calculations and 
simulations~\cite{LeBaTa98,DiBaLe01,LiLo00,LiLo99,Li00}.

The result that the 
non-linear terms omitted within the 
Debye-H\"uckel approximation stabilize a deionized colloidal 
suspensions against a liquid-vapor  phase
separation has also been obtained 
by von Gr\"unberg et al.~\cite{GrRoKl01,DeGr02,KlGr01} and 
Tamashiro and Schiessel~\cite{Ta02} based on the analysis
of the full  non-linear Poisson-Boltzmann
equation inside a Wigner-Seitz cell. The numerical integration of 
the non-linear $PB$ shows that the osmotic pressure
is a  monotonically increasing
function of  colloidal concentration. This means that  
at the level of  $WS$ approximation suspension is thermodynamically
stable. Von Gr\"unberg et al. and Tamashiro and Schiessel,
however, demonstrate that the {\it linearized $PB$ equation} 
leads to the  negative compressibility and the osmotic
pressures for highly charged colloidal 
particles.  This 
erroneously suggests presence of a 
thermodynamic instability. 
Clearly the instability is an artifact
of the linearization.  Furthermore, our calculations show that any
linear theory, which does not take into account the colloidal
charge renormalization, is likely to lead to
an incorrect prediction of a liquid-vapor 
phase separation~\cite{RoHa97,ChLiPe01} in deionized aqueous 
suspensions with monovalent counterions.

It is curious 
that the ``linear'' correlations between the colloids and the counterions, 
responsible for the screening of electrostatic
interactions, are also the ones driving the suspension towards the
phase separation.
On the other hand, the ``non-linear'' correlations
responsible for the counterion condensation
and the colloidal charge renormalization,  stabilize
the suspension against a phase transition.

\section{Polyelectrolyte solutions}

Polyelectrolytes are polymers with ionizable groups 
which have tendency to dissociate in polar solvents~\cite{BaJo96,NeAn02}.  
The good water solubility of polyelectrolytes
is due to large favorable gain in the solvation free energy resulting 
from
hydration of charged monomers and counterions.  
Unlike polyampholytes~\cite{HiJo90,KaKa95,DoRu95,LeBa96}, whose monomers
can  either be cationic (positive) or anionic (negative),
all charged monomers of a polyelectrolyte carry charge of the same sign. 
Depending on the sign of this charge,  polyions are either cationic
or anionic. Over the last few decades polyelectrolytes have
found many industrial applications ranging from water treatment
and oil recovery to detergents and superabsorbants. 
The
biological importance of polyelectrolytes, however, 
has been realized much earlier. 
After all the most important biomolecule, $DNA$, is an 
anionic polyelectrolyte~\cite{Ma78}.

Unlike the simple polymeric fluids, thermodynamics of which is 
fairly well understood,  polyelectrolyte
solutions still remain to large extent enigmatic. The difficulty
in studying polyelectrolytes
resides in the combination of 
polyion flexibility~\cite{GePiVe76,Kh80,StKr95,BaJo96,DoRuOb96,MiHoKr99,LeTh01}, the
long-ranged nature of the Coulomb force, as well as
a large charge and size 
asymmetry between the polyions and the counterions.    
There are,
however, some polyelectrolytes  whose polyions  are 
rigid molecules. This allows to bypass the complications associated
with the statistics of polyion conformations.  
We have already come across this kind of
systems in our exploration  
of charged colloidal suspensions. Many biologically 
relevant polyelectrolytes are also fairly rigid.  Persistence
length (the distance over which polymer can be considered to be
rodlike) for double stranded $DNA$ is on the order of $500$ \AA.
Actin filaments, which are the building blocks of a cytoskeleton,
have persistence length even larger, on the order of microns.  
This should be compared with the Debye length at physiological  
concentrations of $150 \,mM$ of $NaCl$, which is about $8$ \AA.  Clearly
under these conditions the flexibility of the $DNA$ or the 
actin filaments
can be considered an irrelevant perturbation.  
	
The thermodynamics of rodlike polyelectrolytes can be explored
using the same theoretical tools used for spherical colloidal suspensions.
Rodlike molecules can undergo nematic and smectic phase transitions.
For ordered periodic structures,  the Wigner-Seitz cell formalism
can be employed to obtain most of the 
relevant thermodynamics~\cite{FuKaLi51,DeHo01}.
At low volume fraction, when a polyelectrolyte solution is
disordered, this strategy is no longer valid and a different methodology
must be used. This can be constructed along the same lines
taken for colloidal suspensions~\cite{Le96,KuLeBa98}.  
The fundamental role of electrostatic
correlations between the polyions and the counterions appears as
Debye screening of polyion-polyion interactions and the 
renormalization
of a polyion charge. In polyelectrolyte 
literature the polyion-counterion association leading to polyion charge 
renormalization is known as the Manning
condensation~\cite{Ma69,Ma78,Oo71}.  
Here we will show that the Manning condensation
is very similar to the charge renormalization 
found in colloidal suspensions.

First, we shall briefly review Manning's original argument~\cite{Ma69}.
Manning was interested in deriving the limiting (low density) 
laws for polyelectrolyte
solutions, similar to the ones found by Debye and H\"uckel for
simple electrolytes, see Eq.~(\ref{7a}). The salient feature
of the Debye-H\"uckel limiting laws is that they do not
depend on specifics of electrolyte, i.e. size or hydration.  
For example, the osmotic pressure
at low ionic concentrations is found to be a function only of the
ionic charge, temperature, and concentration.  The question
then arises if such a limiting law is also possible for polyelectrolyte
solutions.  The answer to this question is far from obvious.
The strong electrostatic interaction between the polyions and
counterions favors accumulation of counterions in the
vicinity of polyions.  It is, therefore, possible  that even at
very large dilutions the physics of a polyelectrolyte solution 
remains that of a strongly
interacting system  for  which no limiting law should 
be anticipated~\cite{Ma78}.

\subsection{Manning condensation} 
\label{Manning}  
Consider a simple model of a polyelectrolyte solution.  The rodlike
polyions of concentration $\rho_p$, 
idealized as rigid cylinders of length $L$ and radius $a$, 
carrying $Z$ ionized
groups --- each of charge $-q$ uniformly spaced along the major axis
of the cylinder--- inside a uniform dielectric
solvent of constant $\epsilon$.  
The counterions of concentration $\rho_c=Z \rho_p$ 
will be treated as point particles
of charge $q$. For simplicity we will restrict our attention to the
situation in which there is no added salt.  

In the low density
limit we can neglect the discreteness of the polyion charge distribution and
assume a uniform line-charge density,
\begin{equation}
\label{107} 
\lambda_0=-\frac{Zq}{L} \equiv -\frac{q}{b}\;,
\end{equation}
where $b$ is the separation between the successive charged monomers along
the polyelectrolyte chain. 
The bare interaction potential between a long charged cylinder and a counterion
is,
\begin{equation}
\label{108} 
\phi=-\frac{2q\lambda_0}{\epsilon}\ln\left(\frac{r}{r_0}\right)\;,
\end{equation}
where $r_0$ is the arbitrarily  chosen point of zero potential.
A  polyion-counterion two-body partition function is 
\begin{equation}
\label{109} 
\zeta_1=L \int_a^R e^{-\beta q \phi(r)} d^2 r=\pi L r_0 
\frac{(R/r_0)^{(2-2\xi)}-(a/r_0)^{(2-2\xi)}}{1-\xi}\;,
\end{equation}
where $R$ is the cutoff distance at which a counterion cas still be 
considered
to be bound to the polyion.  The Manning parameter is defined as,
\begin{equation}
\label{110} 
\xi=\frac{|q\lambda_0|}{\epsilon k_BT}=\frac{q^2}{\epsilon k_B T b}.
\end{equation}

The integral in Eq.~(\ref{109}) remains finite for all values of $\xi$.
Manning noticed, however, that if the limiting laws exist, the
thermodynamic functions should be independent 
of the polyion diameter.  However, if $a=0$ the integral
in Eq.~(\ref{109}) diverges as $\xi \rightarrow 1^-$.  
Manning interpreted this divergence as an indication of the counterion
condensation.   For values of $\xi>1$ he supposed that $n$
counterions condense onto the polyion, reducing proportionately 
its effective
charge density from $\lambda_0$ to 
\begin{equation}
\label{111} 
\lambda_n=\lambda_0 \frac{Z-n}{Z} \;.
\end{equation}
To find the number of condensed counterions $n$, Manning
postulated that for $\xi>1$ the effective reduced line charge density
\begin{equation}
\label{112} 
\xi_{eff} \equiv \frac{|q\lambda_n|}{\epsilon k_BT}\;
\end{equation}
saturates at one, 
\begin{equation}
\label{113}
\xi_{eff}=1 \;.
\end{equation}
If  the ``renormalized''  $\xi_{eff}$ is used in
Eq.~(\ref{109}), instead of the ``bare'' $\xi$ 
when $\xi>1$, the polyion-counterion partition
function  remains finite. 
Eq.~(\ref{112}) and Eq.~(\ref{113}) determine the number of
condensed counterions to be 
\begin{eqnarray}
\label{114}
n^*=Z\left(1-\frac{1}{\xi}\right) \;\;\;for\;\;\;\xi>1 \nonumber\\
n^*=0\;\;\;\;\;\;\;\;\;\;\;\;\;\;\;\;\;\;for\;\;\;\xi\le 1
\end{eqnarray}

Once $n^*$ is determined, the rest of the thermodynamic functions
can be calculated quite easily.  The nice thing about Manning's
argument is that it is so simple.  On the other hand it 
contains some points which might leave a more mathematically
inclined reader quite disturbed.  Manning relied on the existence
of the limiting laws to establish the limiting law in the first place!
This is a circular logic which is not always guaranteed to work.  
It is interesting  to apply the same argument to the case
of a two-dimensional plasma of particles interacting by a logarithmic
potential~\cite{Le98}, Section \ref{KT}. Consider  
an anion-cation two body partition function Eq.~(\ref{18}).  
Following Manning, lets look at the limit $a \rightarrow 0$.
In this case we find that the integral in Eq.~(\ref{18}) diverges
for the temperatures 
\begin{eqnarray}
\label{115}
T<T_d \equiv\frac{q^2}{2 k_B \epsilon}\;.
\end{eqnarray}
Thus, we might incorrectly conclude that the metal-insulator transition
also happens at $T_d$.  In reality, 
we know that the  Kosterlitz-Thouless  
transition occurs at half this temperature,
\begin{eqnarray}
\label{116}
T=T_{KT}\equiv\frac{q^2}{4 k_B \epsilon} \;.
\end{eqnarray}
For any finite value of the particle diameter, all the thermodynamic functions
are analytic at $T_d$ and the singularities only appear at $T=T_{KT}$.  
Therefore, it seems far from obvious if Manning's argument is valid for
real polyelectrolytes with monomers and counterions of finite diameter.
To explore this in more detail we can appeal to the Debye-H\"uckel-Bjerrum
theory, previously constructed for low dielectric  electrolytes, 
Section  \ref{Bjerrum}.

\subsection{Counterion Association}

We will restrict our attention to the salt-free
infinite dilution limit.  The calculations, however, can
be extended to solutions of 
finite polyelectrolyte concentration as well as to polyelectrolytes
with  salt or even amphiphiles~\cite{KuLeBa98,KuLeBa98a}. As in the case of
colloidal fluids, we would like to trace out the degrees of 
freedom associated with the counterions. This leads to the  
effective many-body interactions between the
polyions.  In the limit of infinite dilution the contribution
from these interactions to the total free energy 
can be neglected.

Consider a dilute polyelectrolyte solution in thermal equilibrium. 
If
we take a snapshot, we will see polyions distributed more
or less 
uniformly throughout the solution, with no specific 
orientation. The counterions, on the other hand, will be clustered
in the vicinity of polyions. This picture suggests that
there are only weak positional correlations between the polyions and
strong positional correlations between the polyions and the
counterions~\cite{Lo94}.
The average distribution of charge around a polyion can, therefore,
be approximated by the Eq.~(\ref{66}).

Suppose we choose a coordinate system so that it is centered
on one of the polyions, with the $z$-axis along the polyion's 
major axis. 
The electrostatic potential for $r<a$ satisfies the Laplace
equation, while for $r \ge a$ it satisfies the Poisson equation
with the charge distribution approximated by the Eq.~(\ref{66}). As in the
case of colloids, we would like to linearize the exponential Boltzmann
factor. This, however, is prohibited by the fact that electrostatic
interactions  are
very strong in the vicinity of the polyion surface.  
Thus, in order to linearize the exponential, the short
distance polyion-counterion correlations must be 
explicitly taken into
account.  In our earlier study of colloidal suspensions this was
done by introducing an effective renormalized charge.  Here
we will take a somewhat different approach.  
The main consequence of short distance electrostatic
correlations is the  polyion-counterion
association. This is very similar to the concept of 
Bjerrum association, which has
proven so successful for  symmetric 1:1 electrolytes.
A polyelectrolyte solution can be thought of as being composed of 
free unassociated counterions of density $\rho_f$,
as well as of  clusters of density $\rho_n$, consisting  of one polyion and 
some number $0 \le n \le Z$ of associated counterions. The polyions
with no associated counterions are treated as $0$-clusters.  
The  particle conservation requires that  
\begin{equation}
\label{67} 
\sum_{n=0}^Z \rho_n = \rho_p\;. 
\end{equation}
and
\begin{equation}
\label{68} 
\rho_f + \sum_{n=0}^Z n \rho_n = Z\rho_p\;. 
\end{equation}
The distribution
of cluster sizes $\{\rho_n\}$ can be determined from the 
equilibrium condition that the Helmholtz free energy be minimum.
Once the clusters are explicitly introduced into the theory,
the exponential factor in Eq.~(\ref{66}) can be linearized,
since at large distances  $\beta q \phi(r)<1$ and at short
distances the ``non-linearities'' are  
accounted for through the cluster formation. 
After linearization, the Poisson-Boltzmann equation reduces to the
Helmholtz equation with
$\kappa$ given by,
\begin{equation}
\label{70}
\kappa=\sqrt{\frac{4 \pi q^2 \rho_f}{k_B T \epsilon}}.
\end{equation}
The linear equation can be easily solved yielding the  
electrostatic potential outside 
a $n$-cluster
\begin{eqnarray}
\label{117}
\phi(r)=\frac{2\lambda_n}{\epsilon} 
\frac{K_0(\kappa r)}{\kappa a K_1(\kappa a)} \;,
\end{eqnarray}
where $\lambda_n$ is the linear cluster charge density of a $n$-cluster
Eq.~(\ref{111}),
and $K_\nu(x)$ is the modified Bessel function of order $\nu$.
For distances $r<a$ the electrostatic potential is found to be
\begin{eqnarray}
\label{118}
\phi(r)=-\frac{2\lambda_n}{\epsilon}\ln(r/a)+\frac{2\lambda_n}{\epsilon} 
\frac{K_0(\kappa a)}{\kappa a K_1(\kappa a)} \;.
\end{eqnarray}
The electrostatic energy due to $n$-cluster-counterion interaction
can be obtained from
Eq.~(\ref{73}) and the electrostatic {\it free} energy follows
from the Debye charging process as in Eq.~(\ref{76}).  In the 
limit of infinite dilution, the electrostatic 
free energy of a $n$-cluster is
\begin{eqnarray}
\label{119}
\beta {\cal F}_n=-\frac{(Z-n)^2}{Z} \xi \left[\frac{2\gamma_E-1}{2}+
\ln\left(\frac{\kappa a}{2}\right)\right]+{\cal O}(\rho_f)\;.
\end{eqnarray}
The electrostatic free energy density $f=F/V$ 
due to all cluster-counterion
interactions is 
\begin{eqnarray}
\label{120}
\beta  f^{pc}=-\frac{ \xi}{Z} \sum_{n=0}^Z  (Z-n)^2 \rho_n 
\left[\frac{2\gamma_E-1}{2}+
\ln\left(\frac{\kappa a}{2}\right)+{\cal O}(\rho_f)\right]\;.
\end{eqnarray}
The cluster-cluster  and the 
counterion-counterion contributions are of higher order in density and
in the limit of infinite dilution can be neglected.   The only 
contributions which must still be taken into account are the entropic free
energy of mixing and the free energy necessary to construct the isolated
clusters. Both of these can be concisely written as   
\begin{equation}
\label{121} 
\beta f^{ent-cl}=
\rho_f \left[\ln( \rho_f \Lambda^3_c)-1\right]+ \sum_n \rho_n \left[ 
\ln \left(\frac {\Lambda_n^{3(n+1)} \rho_n}{\zeta_n}\right)-1\right] \;, 
\end{equation}
where the de Broglie thermal wavelength of a
$n$-clusters is
\begin{equation}
\label{122} 
\Lambda_n=\frac{h}{\sqrt{2 \pi m_n k_B T}} \;, 
\end{equation}
and $m_n$ is the cluster geometric mean mass,
\begin{equation}
\label{123} 
 m_n=(m_p m^n_c)^{\frac{1}{n+1}} \;. 
\end{equation}
The internal partition function of a $n$-cluster is
\begin{equation}
\label{124} 
\zeta_n=\frac{1}{n!}\int \prod_{i=1}^n 
\frac{d^3{\bf r_i}}{\Lambda_n^3}\; e^{-\beta U} \;, 
\end{equation}
were $U$ is the usual Coulomb potential. A suitable cutoff must
be chosen in order to define what constitutes a cluster.  Evaluation
of the integral in Eq.~(\ref{124}) represents a formidable task. 
Fortunately, as we shall see, for polyelectrolyte
solutions at infinite dilution specific knowledge of $\zeta_n$ 
proves unnecessary. 
The total free energy density is then
\begin{equation}
\label{125} 
f(\{\rho_n\})=f^{pc}+f^{ent-cl} \;. 
\end{equation}
To find the equilibrium cluster distribution, this free energy
must be minimized subject to constraints of particle conservation,
Eqs.~(\ref{67}) and  (\ref{68}).  The minimization is equivalent to 
the law
of mass action, 
\begin{equation}
\label{88} 
\mu_n=\mu_0+n\mu_f \;, 
\end{equation}
where the chemical potential of  $n$-clusters is
\begin{equation}
\label{89} 
\mu_n=\frac{\partial f}{\partial \rho_n} \;, 
\end{equation}
and the chemical potential of free ions is
\begin{equation}
\label{90} 
\mu_f=\frac{\partial f}{\partial \rho_f} \;. 
\end{equation}
Substitution of
the total free energy into Eq.~(\ref{88}) leads 
to the $n$-cluster distribution
\begin{equation}
\label{126} 
\rho_n=\zeta_n \rho_0 \rho_f^n 
e^{\beta(\mu_0^{ex}+n\mu_f^{ex}-\mu_n^{ex})} \;,
\end{equation}
which is actually a set of $Z$ coupled algebraic equations.
At large dilutions, the excess  
chemical potential of  $n$-clusters is 
\begin{eqnarray}
\label{127} 
\beta \mu_n^{ex}=-\frac{(Z-n)^2}{Z} \xi \left[\frac{2\gamma_E-1}{2}+
\ln\left(\frac{\kappa a}{2}\right)\right] +{\cal O}(\rho_f)\;, 
\end{eqnarray}
and of free ions is 
\begin{equation}
\label{128} 
\beta \mu_f^{ex}=
-\xi \sum_{n=0}^Z \frac{(Z-n)^2 \rho_n}{2Z\rho_f}+{\cal O}(\rho_f)\;. 
\end{equation}

The  internal partition function of a $n$-cluster is 
independent of density. Recalling that
$\kappa \sim \sqrt{\rho_f}$,  Eq.~(\ref{126}) simplifies to
\begin{equation}
\label{129} 
\rho_n \sim \rho_0 \rho_f^{g(n)}\;, 
\end{equation}
where the exponent is 
\begin{equation}
\label{130} 
g(n)=\frac{\xi}{2Z}n^2-\xi n+ n\;, 
\end{equation}
In the limit of infinite dilution, $\rho_f \rightarrow 0$,
only the cluster of size
\begin{equation}
\label{131} 
n_M=Z\left(1-\frac{1}{\xi}\right)\;, 
\end{equation}
which minimizes $g(n)$ survives.  In this limit the 
cluster size distribution
takes a particularly simple form, 
\begin{equation}
\label{132} 
\rho_n=\rho_p \delta_{n\:n_M}\;. 
\end{equation}
This is precisely the cluster size distribution postulated by Manning based
on his heuristic argument. 
The osmotic pressure of a polyelectrolyte solution can be obtained through 
the Legendre transform of the negative free energy density~\cite{Le98},
\begin{equation}
\label{93} 
P=-f(\rho_f,\{\rho_n\})+\mu_f\rho_f+ \sum_n \mu_n \rho_n\;,
\end{equation}
which leads directly to the Manning limiting law for pressure~\cite{Ma69},
\begin{equation}
\label{133} 
\beta P= \left(1-\frac{1}{2 \xi}\right)Z\rho_p \;\;\;\;for\;\;\;\;\ \xi \le 1 \;.\end{equation}
\begin{equation}
\label{134} 
\beta P= \frac{Z \rho_p}{2 \xi} \;\;\;\;\;\;\;\;\;\;\;\;\;\;\;\;for\;\;\;\;\ \xi > 1 \;. 
\end{equation}

The discontinuity in the slope of pressure as a function of temperature
has provoked a lot of speculation that the Manning condensation is a
real thermodynamic phase transition. Kholodenko and Beyerlein went
even so far as to identify Manning condensation with the 
Kosterlitz-Thouless transition~\cite{KhBe95}. That this is incorrect
follows already from our discussion in Section \ref{Manning}. For
the two dimensional plasma with logarithmic interactions, 
the Kosterlitz-Thouless transition occurs
at half the equivalent Manning temperature. The two phenomena, therefore, 
have nothing in common~\cite{Le98}.  Furthermore, while the 
Kosterlitz-Thouless is a
real thermodynamic phase transition characterized by the diverging
Debye length, for rodlike polyelectrolytes the discontinuity in the slope
of pressure appears only in the double limit $\rho_p \rightarrow 0$,
$L\rightarrow \infty$.  If the order of limits is interchanged, there
is no singularity and no counterion condensation. In addition directly
at the condensation threshold $\xi=1$, the Debye 
length remains finite~\cite{KuLeBa98}.

For polyelectrolyte solutions containing
polyions of {\it finite} size and at {\it non-zero} concentration, 
there is also a polyion-counterion
association~\cite{KuLeBa98}.  
In this case, however, the distribution of cluster
sizes is no longer a delta-function, but rather a bell-shaped curve
centered on the value $n^*$. We find that $n^*$ depends on 
the concentration
of polyelectrolyte  and is somewhat
larger than the limiting Manning value $n_M$. The pressure remains an 
analytic function of $\xi$, showing that for  
real polyelectrolyte solutions, the
counterion condensation is actually a crossover phenomenon, 
very similar
to the  micellar formation in amphiphilic systems~\cite{IsMiNi76}.

\section{Multivalent counterions}

Up to now we have been concentrating our attention on aqueous solutions
with monovalent counterions.  It was already mentioned that in this 
case the correlations between the
condensed counterions can be neglected.  To understand
why, let us  compare the characteristic electrostatic energy of 
a counterion-counterion interaction  
to the characteristic thermal energy $k_B T$,
\begin{equation}
\label{135} 
\Gamma=\frac{\alpha^2 q^2 }{\epsilon d k_B T }\;, 
\end{equation}
where $\alpha$ is the counterion valence and $d$ is the 
average separation 
between the $n$ condensed counterions 
on the surface of a colloidal particle of radius $a$. 
Since $n \pi d^2=4 \pi a^2$,
\begin{equation}
\label{136} 
d=\frac{2 a}{\sqrt n }\;, 
\end{equation}
and the coupling strength becomes,
\begin{equation}
\label{137} 
\Gamma=\frac{\alpha^2 \lambda_B \sqrt n }{2 a }\;. 
\end{equation}
Now, lets consider highly charged latex particles with
$Z=7000$ and
$a=1000$ \AA, in water at room temperature.  
From Eq.~(\ref{65ai}), taking $\chi=15$,  
$Z_{eff}^{sat}=2100$, which means that
$4900$  monovalent $(\alpha=1)$ counterions
are condensed onto the particle. The coupling strength of
the counterion-counterion interaction is then $\Gamma \approx 0.25$, which
clearly shows that the electrostatic interactions between the
condensed counterions are very weak. 
We can make this observation even more general. The high
surface charge concentration $\sigma_m$ encountered in nature 
is on the order
of one elementary charge per  $100$ \AA$^2$.  
Lets suppose that suspension consists of  highly charged 
colloidal particles with  surface charge density $\sigma_m$. 
Clearly this means
that there will be a lot of counterion condensation.  For a
salt-free colloidal suspension containing multivalent counterions,
the number of condensed counterions will be approximately
\begin{equation}
\label{138} 
n^* \approx \frac{Z }{\alpha }\;. 
\end{equation}
The radius of a colloidal particle can be expressed as
\begin{equation}
\label{139} 
a=\sqrt{\frac{Z }{4 \pi \sigma_m }}\;. 
\end{equation}
Substituting Eqs.~(\ref{138}) and (\ref{139}) into Eq.~(\ref{137})
we find that the maximum counterion-counterion coupling 
strength is,
\begin{equation}
\label{140} 
\Gamma_{max} \approx \alpha^{\frac{3}{2}} \lambda_B \sqrt{\pi \sigma_m} \;. 
\end{equation}
For monovalent counterions 
$\Gamma_{max} \approx 1.3$, for divalent 
counterions $\Gamma_{max} \approx   3.6$,
and for trivalent counterions $\Gamma_{max} \approx  6.8$.  Although
$\Gamma_{max}$ is an overestimate, it clearly shows that
for highly charged colloidal particles, correlations between
the condensed multivalent counterions cannot be ignored.

\subsection{Overcharging}

One consequence of strong electrostatic correlations 
is the phenomenon known as the 
``overcharging''~\cite{ToVa80,GoLoHe85,LoSaHe83,MeToLo01,NgSh00b,NgGrSh00b,GrNgSh02,NgGrSh00c,NgGrSh00,TaGr01}.
Overcharging occurs as the result of highly favorable  
gain in electrostatic free energy due to 
strong positional correlations between the 
condensed counterions. 

To understand better how the overcharging of colloidal particles
comes about let us consider a simple case of one
colloidal particle with a uniform surface 
charge $-Zq$ and radius $a$, 
at zero temperature~\cite{MeHoKr00,MeHoKr01}.
The question that we would like to answer is 
how many $\alpha$-valent counterions should be placed
on top of the colloidal particle in order to minimize the
electrostatic energy of the resultant 
polyion-counterion complex?  Naively we might suppose that the number
of condensed counterions should  be such as to neutralize completely
the colloidal charge.  This, indeed, would be the case if
the charge of counterions was uniformly smeared over the
surface of colloid.  In reality, the counterions are discrete entities
and can gain favorable energy by maximizing their separation from
one another. Lets calculate the electrostatic energy
of the polyion-counterion complex,
\begin{equation}
\label{143a} 
E_n=\frac{Z^2 q^2}{2 \epsilon a}-
\frac{Z \alpha n q^2}{\epsilon a}+F^{\alpha \alpha}_n \;.
\end{equation}
The first term is the self energy of a polyion,  the second
term is the electrostatic energy of interaction between the polyion
and $n$ condensed $\alpha$-ions, and  the last term is the
electrostatic energy of repulsion between the condensed counterions.
Now, consider a 
one component plasma of $n$ $\alpha$-ions on the surface of a sphere
of radius $a$ but with a {\it uniform neutralizing background charge} $-\alpha n q$.
The electrostatic energy of this $OCP$ can be expressed
as the sum of contributions arising from the counterion-counterion
interaction,  counterion-background interaction, and the self
energy of the background,
\begin{equation}
\label{143b} 
F^{OCP}_n=F^{\alpha \alpha}_n -\frac{\alpha^2 n^2 q^2}{\epsilon a} + 
\frac{\alpha^2 n^2 q^2}{2\epsilon a}\;.
\end{equation}
Substituting this expression into Eq.~(\ref{143a}), the
electrostatic energy of the polyion-counterion complex simplifies to
\begin{equation}
\label{143c} 
E_n=\frac{(Z-\alpha n)^2 q^2}{2 \epsilon a}+F^{OCP}_n \;.
\end{equation}
For low temperatures, the condensed counterions try
to maximize their separation from one another.  In the planar
geometry the ground state corresponds to a triangular Wigner crystal.
A similar arrangement of counterions will also be found
on the surface of a spherical colloidal particle, up to some topological
defects.  The electrostatic
energy of a planar $OCP$ has been discussed in Section \ref{Cocp}.
For a spherical $OCP$ the 
electrostatic energy at zero temperature is 
\begin{eqnarray}
\label{143d}  
F_{n}^{OCP}=-M\frac{\alpha^2 q^2 n^{3/2} }{ 2 \epsilon a }\;. 
\end{eqnarray}
where $M$ is the Madelung constant.
Because of the topological difference between the plane and the surface of
a sphere, we expect that the Madelung constant 
will not be exactly the same in the two cases.  
The difference, however, should not be
very large as was confirmed in 
recent Monte Carlo simulations~\cite{MeHoKr01}. 
For concreteness,
we shall use $M=1.106$, the value of the planar $OCP$.
 
The effective charge of the polyion-counterion complex in units of $-q$ is
\begin{eqnarray}
\label{143e}  
Z_{eff}=Z-\alpha n^* \;. 
\end{eqnarray}
where $n^*$  is the number of condensed $\alpha$-ions 
which minimize the electrostatic
energy,
\begin{eqnarray}
\label{143f}  
\frac{d E_n}{dn}\Big\arrowvert_{n^*}=0\;. 
\end{eqnarray}
The effective charge is found to be
\begin{eqnarray}
\label{143g}  
Z_{eff}=-\frac{1+\sqrt{1+4 \gamma^2 Z}}{2 \gamma^2}\;, 
\end{eqnarray}
where $\gamma$ is,
\begin{eqnarray}
\label{143h}  
\gamma=\frac{4}{3 M \sqrt \alpha}\;. 
\end{eqnarray}
We see that the effective charge of the polyion-counterion 
complex is inverted
compared to the bare charge $Z$ of the colloidal particle, so that
the complex is overcharged.  For 
highly charged colloids, the effective charge scales as
the square root of the bare charge~\cite{Sh99a,MeHoKr01},
\begin{eqnarray}
\label{143i}  
Z_{eff} \approx -\frac{\sqrt Z}{\gamma}\;. 
\end{eqnarray}

The analysis above was conducted for one colloidal particle at zero
temperature.  For solutions at finite concentration and
temperature we face the same difficulties already encountered
in our earlier discussion of charge renormalization in colloidal
suspensions and polyelectrolyte solutions.  In fact the
problems are even more severe, since for  multivalent
counterions  the Poisson-Boltzmann equation fails completely. 
A logical step is to appeal to the density
functional theory.  The difficulty with the $DFT$ lies in constructing a 
suitable density functional which can take into account the electrostatic
correlations.  For $Z:\alpha $ suspensions without salt, 
such functional
was presented in Section \ref{DFT}.  
It was found  using the $DFT$ and the 
Wigner-Seitz cell formalism~\cite{Gr91},
that the effective charge 
$Z_{eff}$, as the function
of the bare charge $Z$,
after reaching the  maximum decreases, 
asymptotically going to zero 
as $Z \rightarrow \infty$.  This behavior  is in striking contrast to the
saturation of the effective charge 
predicted by the Poisson-Boltzmann theory~\cite{AlChGr84}.
Unfortunately it is difficult to construct
a suitable density functional for suspensions containing salt.
One alternative is to use the integral equations.  
The numerical complexity of these theories, however, 
tends to obscure the
essential physics of the problem. 
Below we shall present a simple phenomenological model of 
overcharging.  Our goal is not the quantitative accuracy, but
rather the physical insight into the mechanisms leading to the
overcharging in the polyelectrolyte solutions. 

\subsection{Overcharging in electrolyte solutions}

Consider a dilute colloidal suspension containing 
a monovalent salt at concentration
$c$, and an $\alpha$-valent salt at  concentration $c_\alpha$. 
In aqueous solution the monovalent salt dissociates producing $1:1$ 
electrolyte,
while the  $\alpha$-valent salt dissociates into  $\alpha:1$ electrolyte. 
The inverse Debye length is  
\begin{equation}
\label{159} 
\xi_D^{-1}=\kappa=\sqrt{8 \pi \lambda_B I}\;,
\end{equation}
where the ionic strength is 
\begin{equation}
\label{160}
I=\frac{1}{2}\left(\alpha^2 c_\alpha +\alpha c_\alpha+2c \right)\;.
\end{equation}
For simplicity we shall restrict our attention
to suspensions with monovalent salt near 
physiological concentrations ---  $150mM$  of $NaCl$.  
Under these conditions
the Debye length is around $8$ \AA,
and the interactions between the
colloidal particles  can be completely neglected.
Furthermore, since
the electrostatic attraction between the highly charged $\alpha$-ions
and colloids is so much stronger than the interaction
between the monovalent counterions and colloids, the effective
charge of the polyion-counterion complex   
is completely determined by the number of condensed
$\alpha$-ions.

We will define the counterions as free  (not-associated) if they
are farther than some distance $\delta$ from the colloidal surface.
The  ``agglomerate''  is then defined as the  polyion with
its $\delta$-sheath of surrounding counterions. We note, however, 
that not
all of the $\alpha$-ions  in the agglomerate are actually
associated with the polyion.  The reason for this
is that many of these ions  have sufficient kinetic (thermal) energy to 
leave the vicinity of colloidal particle. Only the counterions
which have the total (kinetic plus potential) energy  
less than zero can be considered  bound to the polyion. 
Unfortunately, it is not 
easy to come up with a practical implementation of this energetic
criterion, except within a Molecular Dynamics Simulation.  
On the other hand, a simple 
geometric criterion based on the polyion-counterion
separation is easily implemented.  
Care, however, must be taken not
to count all of the $\alpha$-ions inside the agglomerate as belonging
to the polyion-counterion complex.  Bellow we shall see how this
can be accomplished.
 
We shall take $\delta$ to be on the order of few angstroms,
corresponding to the radius of a hydrated ion, $\delta \approx 2$ \AA.
Since the agglomerate
is in contact with the bulk, its size is determined
by the minimum of the grand potential function
\begin{equation}
\label{161} 
\Omega(n)=F(n)-n \mu_\alpha\;,
\end{equation}
where $F(n)$ is the Helmholtz free energy of the agglomerate
and $\mu_\alpha$ is the chemical potential of  $\alpha$-ions inside
the sheath. In thermal equilibrium, 
the chemical potential of $\alpha$-ions  inside the sheath
equals to the chemical potential of $\alpha$-ions in the bulk
of the suspension. For low bulk concentrations, $\mu_\alpha$
can be approximated
by the chemical potential of an ideal gas,
\begin{equation}
\label{162} 
\mu_\alpha=\ln (c_\alpha \Lambda^3)\;.
\end{equation}
The Helmholtz free energy of an agglomerate is then
\begin{equation}
\label{163} 
F_n=E_n+F_n^{solv}+F_n^{ent}\;.
\end{equation}
The electrostatic free energy of an isolated polyion-counterion
agglomerate $E_n$ is given by the Eq.~(\ref{143c}),  
the solvation energy that the agglomerate gains  
when placed in an 
ionic environment is,
\begin{equation}
\label{164}
F_n^{solv}=-\frac{(Z-\alpha n)^2 q^2 \kappa a}{2 \epsilon a (1+\kappa a)}\;,
\end{equation}
see  Eq.~(\ref{77}), and
the entropic 
energy of counterions inside the sheath is
\begin{equation}
\label{165}
F_n^{ent}=k_B T [n\ln(\rho_n \Lambda^3)-n]\;.
\end{equation}
where,
\begin{equation}
\label{166}
\rho_n=\frac{n}{4 \pi a^2 \delta}\;.
\end{equation}
For high valence counterions (strong-coupling limit) 
the free energy of the $OCP$ 
can be approximated by that of a Wigner crystal, see Section \ref{OCP}.  
If more
accuracy is needed, one can use the extrapolation formulas
based on Monte Carlo simulations~\cite{GaChCh79}.  Here, however, 
we shall content ourselves with the simplest approximation.
The grand potential of an agglomerate is,
\begin{equation}
\label{167} 
\beta \Omega(Z,n)=\frac{(Z-\alpha n)^2 \; \lambda_B}{2 a (1+\kappa a)}
-M\frac{\alpha^2 \lambda_B  n^{3/2} }{ 2 a }+n\ln(\rho_n/c_\alpha)-n\;.
\end{equation}
The number of  $\alpha$-ions $n^*$ inside an agglomerate 
is determined from the minimum
of the grand potential, 
\begin{equation}
\label{168} 
\frac{\partial \Omega(Z,n)}{\partial n}\Big\arrowvert_{n^*}=0\;.
\end{equation}
From the  previous discussion recall that  
it is incorrect to associate the effective charge of
the polyion-counterion complex with the
value of $n^*$, i.e. $Z_{eff} \ne Z-\alpha n^*$.  The reason 
for this is that not all of the $\alpha$-ions inside the 
$\delta$-sheath are actually bound to the polyion. The
real number of condensed counterions is $n^*-n^*_o$, where
the overestimate $n^*_o$ can be obtained    
by considering the number of $\alpha$-ion  
within the distance $\delta$ from the surface of a ``neutral''  
polyion, $Z=0$
\begin{equation}
\label{169} 
\frac{\partial \Omega(0,n)}{\partial n}\Big\arrowvert_{n^*_o}=0\;.
\end{equation}
The effective charge of the polyion-$\alpha$-ion complex is then
\begin{equation}
\label{170} 
Z_{eff}=Z-\alpha (n^*-n^*_o)\;.
\end{equation}
If a small  electric field is applied to the suspension, it is the $Z_{eff}$
which will determines the electrophoretic 
mobility of colloidal particles~\cite{TaGr01,FeFeNi00,Go00,FeFeNi01}.

\begin{figure}  
\begin{center}
\includegraphics[width=8cm,angle=270]{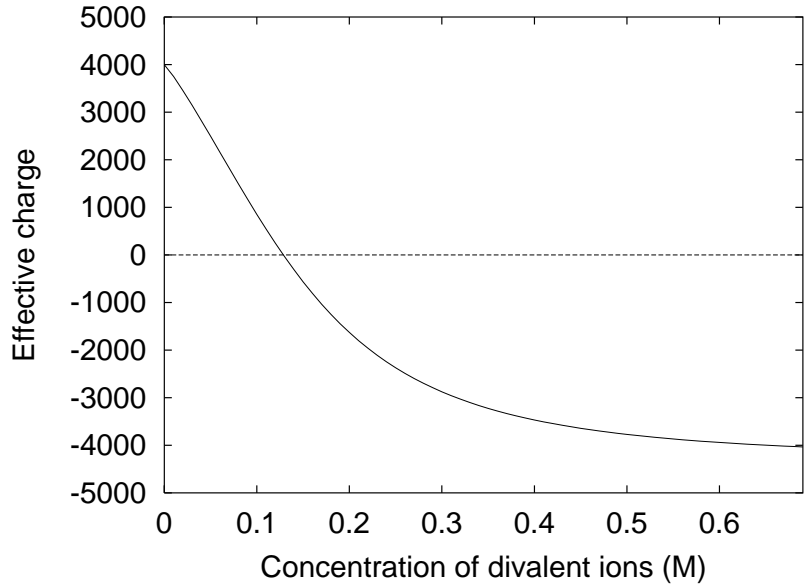}
\end{center}
\caption{The effective (renormalized) charge of colloidal particles
of $Z=4000$, $a=1000$ \AA $\,$ inside a suspension containing monovalent salt
at physiological concentration of $0.15 M$, as a function of
concentration of {\it divalent} counterions.}
\label{Fig3}
\end{figure}

\begin{figure}  
\begin{center}
\includegraphics[width=8cm,angle=270]{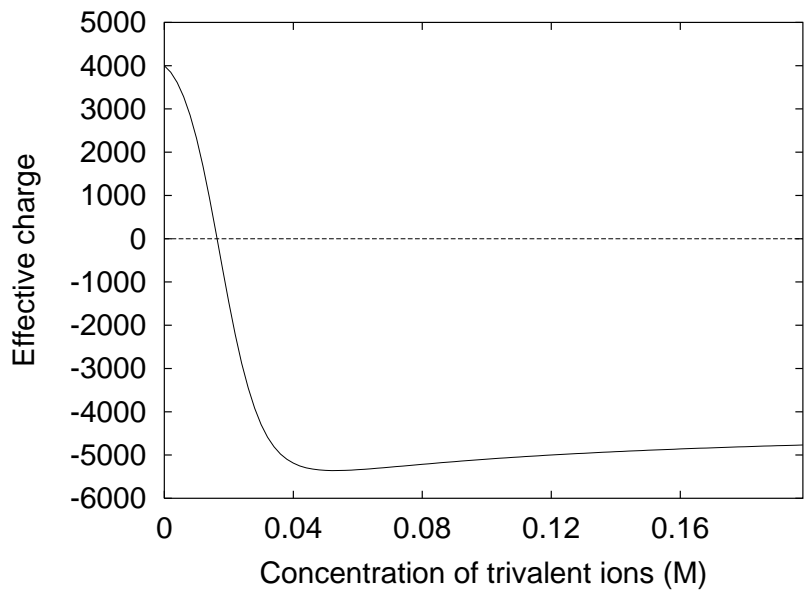}
\end{center}
\caption{The effective (renormalized) charge of colloidal particles
of $Z=4000$, $a=1000$ \AA $\,$ inside a suspension containing monovalent salt
at physiological concentration of $0.15 M$, as a function of
concentration of {\it trivalent} counterions.}
\label{Fig4}
\end{figure}

\begin{figure}  
\begin{center}
\includegraphics[width=8cm,angle=270]{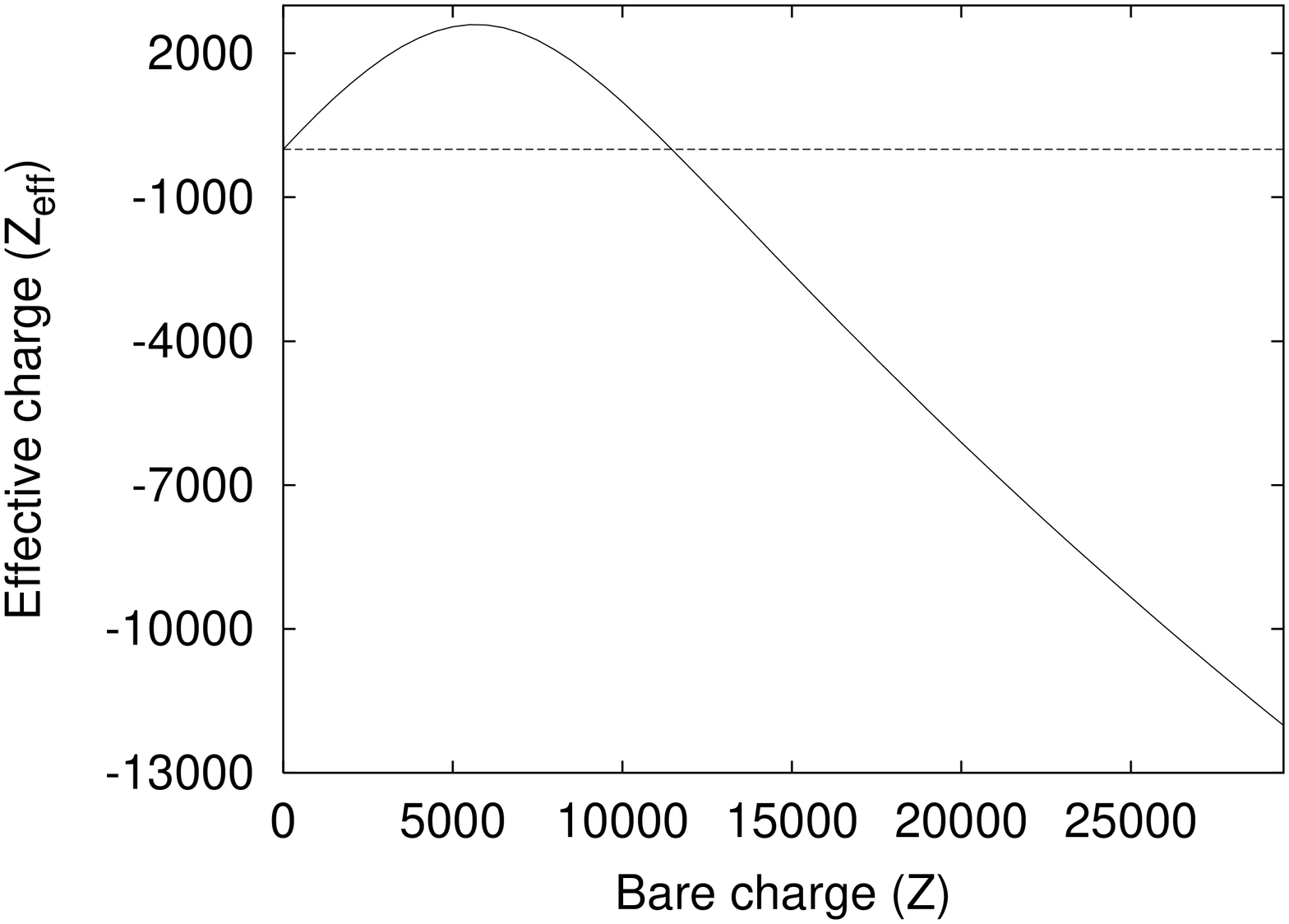}
\end{center}
\caption{The effective (renormalized) charge of a colloidal particle
with $a=1000$ \AA $\,$ as a function of the bare charge $Z$, 
inside a suspension containing monovalent salt
at physiological concentration of $c=0.15 M$ and the trivalent ions
at $c_3=0.01 M$.}
\label{Fig5}
\end{figure}

\begin{figure}  
\begin{center}
\includegraphics[width=8cm,angle=270]{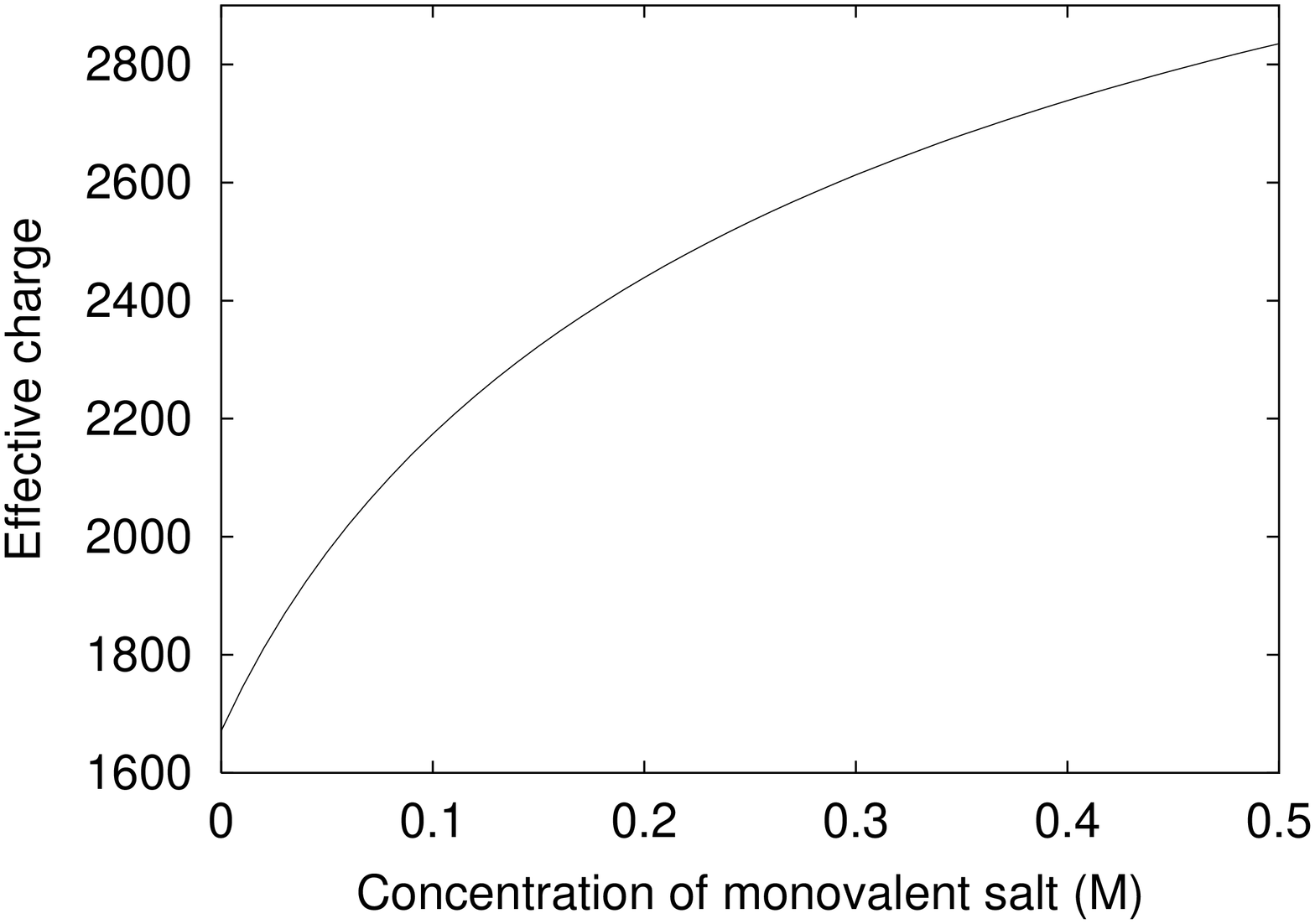}
\end{center}
\caption{The effective (renormalized) charge of colloidal particles
with $Z=4000$ and  $a=1000$ \AA $\,$ inside suspension containing
trivalent counterions with $c_3=0.01M$,  as a 
function of concentration of monovalent salt $c$.}
\label{Fig6}
\end{figure}

\begin{figure}  
\begin{center}
\includegraphics[width=8cm,angle=270]{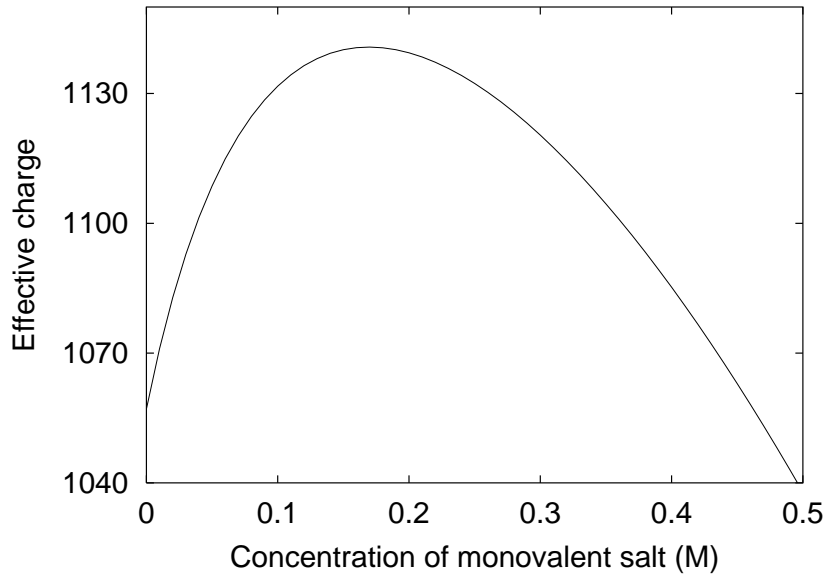}
\end{center}
\caption{The effective (renormalized) charge of colloidal particles
with $Z=4000$ and  $a=1000$ \AA $\,$ inside the suspension containing
trivalent counterions with $c_3=0.0135M$,  as a 
function of concentration of monovalent salt $c$.}
\label{Fig7}
\end{figure}

In Figs. \ref{Fig3} and \ref{Fig4}
we present the effective colloidal charge as
a function of concentration of divalent and trivalent counterions,
for suspension containing colloidal particles
with $Z=4000$ and $a=1000$ \AA.
It is curious to note  the appearance of a  minimum in the effective
charge as a function of trivalent ion concentration.  
In Fig. \ref{Fig5} we present a plot of  
the effective charge as a function
of  bare charge for colloids with $a=1000$ \AA $\,$, 
in a suspension containing  monovalent salt
at concentration $c=0.15 M$ and trivalent counterions
at $c_3=0.01M$.
We note that unlike suspensions containing 
monovalent counterions, the effective charge does not saturate,
instead it reaches  the maximum value 
and then falls off sharply. For colloids
with $Z \approx 11500$, the bare colloidal charge is completely
neutralized by the $\alpha$-ion condensation. 
In Fig. \ref{Fig6} we show the dependence of $Z_{eff}$
on the amount of monovalent salt.  For small concentrations
of $\alpha$-ions,
the effective charge of the complex 
is found to increase with the concentration of monovalent salt,
asymptotically approaching the bare value.
However, when the concentration of multivalent ions reaches the critical
value, there is a qualitative change of behavior. At this point the
effective charge is no longer a monotonically increasing function 
of the monovalent 
salt concentration.  Instead after reaching the  maximum, $Z_{eff}$ begins
to decline, eventually going through the isolectric point $(Z_{eff}=0)$ and
charge inversion, Fig. \ref{Fig7}.  

We have considered only the simplest form of overcharging
involving colloids and multivalent microions.  There is a number
of interesting variations. Recent developments in the field of
gene therapy require construction of  safe and efficient
trans-cellular gene delivery systems~\cite{Fr97,Fe97,FeRi89}. 
Both the DNA and the phospholipids, which are the main constitutive
component of a cellular membrane, are negatively charged.  
There is, therefore,
a strong electrostatic repulsion between the DNA
and the cellular membrane.  This repulsion inhibits
the transfection of naked DNA into the cells.  
Furthermore,  {\it in vivo} the unprotected DNA is rapidly
degraded by the nucleases present in plasma~\cite{HoMuAn98}. 

Much of the effort of gene therapy 
has been concentrated
on viral transfection. A retrovirus (virus which incorporates
its genetic material into the host genome) or adenovirus (which does 
not) has its genetic
material removed and substituted by the gene that needs to
be replicated. The modified viruses are then
made to infect the cells, thus effectuating transfer of the 
genetic
material.  There are, unfortunately, a number of serious
complications involved with this procedure.  These range
from a strong immunological response of an organism against 
the infecting virus, to 
potentially deadly consequences arising from the 
recombinant viral structures~\cite{HoMuAn98}.  
All these factors have
lead to attempts to develop  non-viral transfection methods. 
One of the promising approaches relies on formation of
the DNA-cationic lipid complexes, or lipoplexes for short.  
The hydrophobic interaction between the lipid tails,
in addition to the cationic charge of their  head groups,
favors their agglomeration in the vicinity of polyions.
For sufficiently long hydrocarbon chains, the gain
in the hydrophobic energy resulting from the lipid condensation
onto the DNA is sufficient to lead to the charge reversal of
a lipoplex~\cite{KuLeBa99}.  This happens even though the charged head group
of the lipid is monovalent.  The overcharging
in this case is the consequence of hydrophobicity of lipid molecules and
is not the result of electrostatic correlations.  

A different method for gene transfection  uses
multivalent counterions such as $Ca^{++}$.  
The efficiency of $Ca^{++}$ as a
transfection agent might be due to its ability to neutralize or
even invert the helical DNA charge without making it collapse.
In this respect it is  quite different from the
transition metal ions such as $Mn^{++}$ and $Cd^{++}$
which neutralize the DNA charge,
but also lead to its condensation~\cite{KoLe99}.
Formation of a neutral or overcharged 
DNA-$Ca^{++}$ complex allows the DNA to come into a close
proximity of the cellular membrane.  
Transfection might then be able to proceed through a 
reptation-like motion of the DNA through a cellular pore.  

A very curious form of 
overcharging is found to occur in cellular compaction of the
DNA. Virtually all of the DNA
in  nuclei of eukaryotic cells 
exists as a highly organized nucleoprotein fiber
called chromatin.  At the lowest level of the chromatin
hierarchical structure is a nucleosome. Nucleosome  
resembles
a thread wound around a cylindrical spool.  The thread 
is the DNA molecule, while
the spool is an
octameric protein composed of 
eight smaller proteins called histones.  Each octamer has
$220$ basic residues of which $148$ are on the surface of the
protein and are exposed to the solvent~\cite{KhDrSi97}.  
The rest are inside the protein
core and are unlikely to be ionized. 
The maximum charge of an octamer is, therefore, $Q_{oc}^{max}=+220e$, 
but  is likely to be
significantly lower than this value under the normal 
physiological conditions. 
Each octamer is encircled
by 1.8 turns ($147$ base pairs)
of the DNA thread, carrying a net charge of $-294e$. 
A nucleosome is, therefore, strongly overcharged by 
the associated DNA~\cite{GrNgSh02,AnJo00,PaBrGe99,MaJePi99}.

It is interesting to speculate how the nature is using this
overcharging in the overall organization of the chromatin. 
The degree of compaction achieved inside the nucleus is quite
astonishing.  The total length of the DNA in the nucleus of a 
human cell is about $3.3$ billion base pairs. If extended it would
be  more than $1m$ long.  Yet,
it is compacted into a  nucleus of diameter of $10 \mu m$! 
Furthermore, all this compaction is done in such a way that
the DNA is easily  transcribed  and replicated.  
This  is, indeed, an astonishing feat of chemical engineering! 
       
\section{Like-charge attraction}

\subsection{Confined suspensions}

One of the most curious phenomena which has produced
much healthy debate in the condensed matter community
is the appearance, under some circumstances, 
of attraction between like-charged macromolecules.
The first observation that something might be missing in the
traditional $DLVO$ theory of colloidal stability came
from the experimental observations of Ise et al. of void structures
inside the  highly deionized suspensions~\cite{SoIs84}.
The finding of voids and clusters  has 
lead Sogami and Ise to propose
a modification of the  $DLVO$
pair potential.  In its stead they suggested a new potential
obtained from the  considerations of the Gibbs free energy of
suspension. 
The Sogami-Ise
potential has created a lot of stimulating controversy,
which is still in progress.  The problem has been
re-analyzed by Overbeek~\cite{Ov87}, who has argued that the Sogami-Ise
theory contains a basic thermodynamic inconsistency, which
when resolved leads to the usual $DLVO$
potential.  This, however, has failed to settle the issue
and a number of  papers are still being published
on the thermodynamics of suspensions whose particles 
interact by the Sogami-Ise potential.  

A recently developed  Digital Video 
Microscopy $(DVM)$ has the potential
of putting an end to this long standing debate. 
$DVM$ provides the possibility of explicitly measuring the
interaction potential between two macromolecules in suspension.
For dilute bulk suspensions, the interaction potential between
two spherical colloidal particles was found to be completely 
consistent with
the $DLVO$ theory~\cite{CrGr94}.  
A  surprising result appeared, however, when a
highly deionized  suspension was 
confined between the glass plates~\cite{CrGr94,KeFr94,GrCr00,Gr00}.  
It was discovered that when
particles were close to a wall, the pair interaction
potential developed a strong attractive component. 
The attraction was  quite long ranged, 
comparable in its extent to the diameter of colloidal particle.

It is well known that the glass-water interface
has a significant negative charge due to dissociation
of silanol groups.  This charge density is comparable to that of
colloidal particles.   
Furthermore, the absence of multivalent ions signifies that electrostatic 
correlations play only a marginal role. Under these
conditions the  Poisson-Boltzmann
theory should work quite well.  Indeed, the first numerical solution of
the $PB$ equation  for two macromolecules inside a cylindrical 
pore has found an attraction between the two polyions~\cite{BoSh98}.  
The triumph of the $PB$ theory was, however,
short lived.   Soon afterwards the mathematical  
proofs were published demonstrating that the
$PB$ equation is incapable of producing attraction between like-charged
particles in a confined geometry~\cite{Ne99,SaCh99,Tr00,TrRa99}.  
The numerical calculation was flawed, 
a consequence of the intrinsic difficulty
of solving numerically
non-linear partial differential equations 
with complicated boundary conditions.  Indeed, it is extremely difficult
to understand how a boundary can possibly change 
the interaction potential from repulsive to attractive.  This is not
to say that the presence of a boundary does not have a 
profound effect on the interactions.
Consider one colloidal particle  inside an electrolyte 
solution  at fixed distance from a wall or an interface.  
In general, the interface
is characterized by a dielectric discontinuity.  For the moment, however,
lets ignore this complication and suppose that the two sides 
have exactly the same dielectric constant.  The interface,
then, separates two half-spaces,  one containing electrolyte
and another electrolyte-free.  The fact that all the charges are confined
to one half-space strongly modifies the distribution 
of ions around the colloid.
While  the ionic cloud is spherically symmetric for colloidal particles
far from the interface (in the bulk), it develops a strong
asymmetry, which results in a net dipole moment of the colloid-electrolyte
system.   This dipole  produces an electric field,
\begin{equation}
\label{171} 
E \sim \frac{1}{r^3}\;.
\end{equation}
which, because of the
interface, can not be screened.  If there are two macromolecules
separated by a distance $r$ along the interface,
the electric field produced by one macromolecule
interacts with the dipole moment ${\bold p}$ of 
the charge distribution
induced by the second
macromolecule. This leads to the effective interaction potential
which is repulsive and falls off as
\begin{equation}
\label{172} 
w(r) \approx {\bold p}\cdot {\bold E} \sim \frac{1}{r^3}\;,
\end{equation}
along the direction of an
interface~\cite{Ja82a,Ja82b,GoHa98,GoHa99,AlHaMe01,HaLo00}.  
This is exactly the same kind
of effective interaction found for 
2d $OCP$, as seen in  Section. \ref{Cocp}.

It is clear that existence of an interface or a wall 
strongly modifies the interaction
potential between two macromolecules.  Instead of an exponential 
screening found in the  bulk, the
interactions  along the interface
fall-off algebraically as $1/r^3$. The potential, however,
is still repulsive, and it is difficult to see how anything
can modify this conclusion at the level of electrostatics.

An interesting suggestion, which seems to account for the
apparent attraction between the like-charged colloids
near a wall, has been recently advanced by Squires and 
Brenner~\cite{SqBr01}.
These authors attributed the attraction to 
non-equilibrium hydrodynamic flows which were not
properly accounted for in experiment.  

While the hydrodynamics
seems to be able to explain the 
apparent attraction between the 
colloids near
a  wall, it is not sufficient to explain the results of 
experiments in which colloids are sandwiched between two glass plates, 
since in this
geometry the hydrodynamic attraction mediated by one wall is 
suppressed by the second wall~\cite{Gr02}. Furthermore,  hydrodynamics 
does not help to understand the long-lived metastable crystalline
structures observed by Grier {\it et al.}~\cite{Gr00,LaGr96} 
when a low density suspension is compressed against 
a glass plate.  If the interactions between 
particles are effectively
repulsive, once the constraint is removed the crystals should
melt within seconds. Instead some crystalline regions are 
found to survive for as long as an hour. What is even more surprising
is that the  
crystallites are actually three dimensional, extending 
far beyond the region were the pairwise surface-mediated 
attraction is found.
This phenomenon is  very similar to the 
voids observed by Ise {\it et al.}.

The nature of  confinement-induced attraction between the like-charged
particles remains an open question. However, we must stress again 
that concentration on  pair potentials when studying  
{\it thermodynamic stability} of colloidal suspensions 
is a serious oversimplification.  
The $DLVO$ theory was proposed
as an indicator of {\it dynamical stability}
against flocculation, 
driven by short-ranged van der Waals  forces.
If the equilibrium structure 
of a colloidal suspension is in question, inter-colloid  
pair potential is not sufficient and the
full free energy must be considered. We have
already seen in Section \ref{Cfluid} 
that a large gain in solvation free energy obtained
from the polyion-counterion interactions, strongly
favors the phase separation of suspension into the coexisting high
and low density phases.  This tendency is opposed 
by the counterion condensation, which renormalizes the effective
colloidal charge. The theory presented in Section \ref{Cfluid} 
suggest that colloidal suspensions
should phase separate when  ${\cal C} > 15.2$. 
The counterion condensation, however, prevents ${\cal C}$ from 
reaching this threshold.  Nevertheless for highly charged
colloidal particles, ${\cal C}$ can come very
close to the critical value, Eq.~(\ref{65ai}). 
This suggests that
a deionized suspension of
highly charged particles might actually be 
very close to criticality. 
This regime will be characterized by strong density
fluctuations, which might appear as
coexisting domains of voids and crystallites. 

\subsection{Correlation-induced attraction}

DNA in aqueous solution is highly ionized due to dissociation
of  phosphate groups.  This ionization results in one 
of the highest charge densities found in nature, one electron
charge every 1.7 \AA. In spite of this huge charge concentration,
over a meter DNA is packed into a nucleus of few micrometers.
This efficient compaction is accomplished with the 
help of cationic proteins.
The  bacteriophages (viruses that infect bacteria) also use multivalent
cations to package their DNA. Thus, the $T7$
bacteriophage head is $10^{-4}$ times smaller than the unpacked form 
of its DNA~\cite{GoSc76}.
Furthermore,  it is found that if the multivalent polyamines, 
known to exist in the host bacteria, are added 
to an {\it in vitro } solution containing DNA, 
the chains condense forming toroids very similar in size and
shape to the ones found {\it in vivo}~\cite{Bl91,Bl97}. 
To produce condensation, 
multivalent counterions must somehow induce attraction
between the different parts of the 
DNA~\cite{StPoRa98,PoPa98,GoKaLi99,SoCr00,ArAn02}.  
The toroidal geometry is the result of the high intrinsic rigidity
of the DNA molecule,  which has the persistence length of $\xi_p=500$ \AA.
If the compaction is done in such a way that the local radius 
of curvature,  $r_c$, exceeds $\xi_p/2 \pi$ the cost in 
elastic energy will be prohibitively high.  
Requirement that $r_c>\xi_p/2 \pi$, therefore,
results in  toroidal or spool-like condensates~\cite{Od98,PaHaGe98}.

In eukaryotic cells the
cytosol is traversed by a complicated network of  
microfilaments which
are made of a protein called F-actin~\cite{Ta96,TaJa96}. 
In spite of its high
negative charge density F-actin, in the presence of multivalent counterions, 
agglomerates forming a network of bundles~\cite{BoBrGe02}.
Addition of monovalent salt  screens the electrostatic
interactions
and re-dissolves the bundles~\cite{TaSzJa97}.   
What is the action of multivalent counterions which induces
attraction between the like-charged 
macromolecules~\cite{StRo90,RoBl96,GrMaBr97,GrMaBr97,HaLi97,LeArSt99,ArStLe99,Sh99,KoLe99,KoLe00a,AlAmLo98,KaGo99,GrBePi98,LaLePi00,SoCr99,Sc99,Sc01}?  To understand
this we shall look at some very 
simple models~\cite{ArStLe99}. 

Consider first two parallel polyions separated by a distance $d$ 
inside a dilute 
solution containing  $\alpha$-valent ions.  
The polyions will be idealized as rigid lines of charge of length
$L=Zb$.  Each line has $Z$ monomers of charge  $-q$  
spaced uniformly along the 
chain.  The solvent is  a uniform medium of dielectric
constant $\epsilon$.  It is convenient to define the reduced
polyion charge density as $\xi=q^2/\epsilon k_BTb$ \cite{Ma69}.
A simple Manning argument then suggests that for $\xi> 1/\alpha$ 
\begin{equation}
\label{172a} 
 n_c=\frac{Z}{\alpha}\left(1-\frac{1}{\alpha \xi}\right)\;
\end{equation}
$\alpha$-ions  condense onto each polyion.  This is the lower
bound on condensation, since the Manning argument
does not take into account the
correlations between the condensed counterions.  Nevertheless,
even this simple estimate suggest that $88\%$ of the DNA's charge
should be neutralized by the divalent counterions.

\begin{figure}  
\begin{center}
\includegraphics[width=8cm,angle=270]{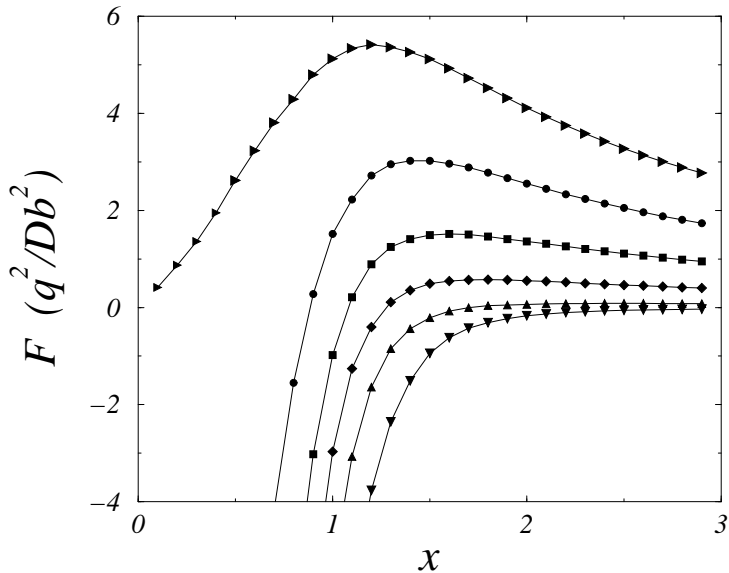}
\end{center}
\caption{Force versus distance between polyions for $Z=20$, $\alpha=2$,
$\xi=2.283$ (corresponding to polymethacrylate) and $n=5,\ldots,10$ (from top
to bottom) in the Monte Carlo simulation~\cite{ArStLe99}. Positive force
signifies repulsion  between the complexes,
while the negative force implies existence of attraction.}
\label{Fig8}
\end{figure}

The associated counterions are free to move along the length of the
DNA.  We shall suppose that the only effect of condensation
is the local renormalization of the monomeric charge  
from $-q$ to $(-1+\alpha)q$.  
Lets define the occupation variables 
$\sigma_{ij}$,
with $i=1, 2, ...,Z$ and $j=1,2$, in such a way that $\sigma_{ij}=1$,
if a counterion is condensed at $i$'th monomer of the $j$'th
polyion, and $\sigma_{ij}=0$ otherwise.
The occupation variables obey
the constraint 
\begin{equation}
\label{173} 
\sum_{i=1}^Z \sigma_{i1} = \sum_{i=1}^Z \sigma_{i2} = n\;,
\end{equation}
where $n$ is the number of condensed $\alpha$-ions.
The interaction energy between the two polyions is
\begin{equation}
\label{174}
 H=\frac{1}{2 \epsilon}\sum_{i,i'=1}^{Z} \sum_{j,j'=1}^{2}
\frac{q^2(1-\alpha\sigma_{ij})(1-\alpha\sigma_{i'j'})}
{r(i,j;i',j')} \;,
\end{equation}
where the sum is restricted to $(i,j) \neq (i',j')$, and
\begin{equation}
\label{175} 
r(i,j;i',j')=b\sqrt{|i-i'|^2 + (1-\delta_{jj'})x^2}\;
\end{equation}
is the distance between the
monomers located at $(i,j)$ and $(i',j')$. 
$\delta_{jj'}$ is the Kronecker delta, and $x=d/b$.
The partition function is
\begin{equation}
Q={\sum_{\{\sigma_{ij}\}}}^\prime\exp(-\beta H) \;,
\label{176}
\end{equation}
where the prime indicates that the trace is done under the constraint
of Eq.~(\ref{173}).  The force between the  two polyions is
\begin{equation}
F=\frac{1}{b \beta}\frac{\partial \ln Q}{\partial x}.
\label{177}
\end{equation}
This model is so simple that for polyions with not too high
values of $Z$, the partition function can  be solved
explicitly~\cite{ArStLe99}.  
For larger $Z's$ the model can be easily simulated.
In Fig. \ref{Fig8} we present the force as a function of separation
for two polyions with $Z=20$ and $n$ condensed divalent
counterions.  
We see that in spite of the net like-charge, the two polyion-counterion
complexes can attract each other at sufficiently small separations.
Furthermore, we find that a critical number $n_l =  Z/2\alpha$ 
of condensed $\alpha$-ions is necessary for the attraction to appear
and that for monovalent counterions the interaction 
is always repulsive.

What is the nature of this attraction?  
To understand this lets consider the limit of small 
separations between the polyions. 
The configurations which dominate the partition function 
for $x \rightarrow 0$ are  the ones in which a condensed counterions on one
polyion face the bare monomeric charges of the second polyion.  In this limit
the positions of condensed counterions on the two
polyions become strongly correlated. 
The electrostatic energy of such configurations is
\begin{equation}
E \approx  \frac{2 n (1-\alpha)q^2}{\epsilon b x}+
\frac{(Z-2n) q^2}{\epsilon b x} \;,
\label{178}
\end{equation}
where  $n$ is the number of condensed counterions.
The first term of Eq.~(\ref{178}) 
is the electrostatic  
energy due to attractive interaction between $n$ condensed
counterions and the bare monomers, while the second
term is the repulsive interaction between $Z-2n$ uncompensated
monomeric charges.
We see that {\it if there are  $n > Z/2\alpha$ condensed counterions} 
the first term of Eq.~(\ref{178}) dominates
and the force becomes attractive 
for sufficiently short separations.  
For $n < Z/2\alpha$ the force is repulsive for all temperatures.  
This is the general mechanism 
of attraction between the two polyion-counterion complexes 
at any finite temperature. 
To minimize the electrostatic {\it free} energy, 
the positions of condensed  counterions
on the  two polyions become correlated.
For sufficiently short separations and 
number of condensed counterions
exceeding the critical threshold $n_l$,  
the correlation-induced
attraction  dominates the 
monopolar repulsion between
the two complexes.
 
The strength of the counterion correlations increase
with a decrease in temperature.  The state of maximum correlation 
is at $T=0$.  This, however, does not imply that
the attraction between two polyions is  
maximum at zero temperature~\cite{StLeAr02}.
The reason for this is that the ground state
configuration, which corresponds to the lowest electrostatic energy, 
is not, in general, the configuration which 
maximizes the attractive force.
For low temperatures and short separations  
there are configurations which have force more attractive than
the force in the ground state.  For example, consider
a polyion with $Z=6$ and $n=3$ condensed divalent counterions.  
The ground state
corresponds to a staggered arrangement of  
counterions on the two polyions.  The counterions 
on one polyion form the pattern $+-+-+-$ while on the
second polyion they form a complimentary pattern $-+-+-+$.
This  leads to attraction
between the complexes.  However, it is easy to see
that the configuration in which the counterions on the two
polyions form the patterns $+++---$ and $---+++$, has
larger electrostatic energy, but leads to stronger attraction.   
At finite temperature the total force, being a weighted
mean of forces associated with all the configurations can, 
therefore, become
more attractive than the force at $T=0$.  This curious behavior, 
however, is confined to very small 
separations between the polyions.  At larger
distances the modulus of 
the attractive force is a monotonically decreasing function of 
temperature.

If the number of condensed counterions exceeds
\begin{equation}
n_u=\frac{Z(\alpha +1)}{2\alpha} \;,
\label{178a}
\end{equation}
the interaction between the two lines of charge becomes, once again, 
repulsive.  The $n_u$ corresponds to an overcharged
configuration in which each polyion has the effective charge
\begin{equation}
Z_{eff}^u=Z-\alpha n_u= -\frac{(\alpha-1)}{2}Z \;.
\label{178b}
\end{equation}
It is curious  that exactly this kind of reentrant behavior is
observed for the DNA condensates~\cite{PeDuDo96,PeLiSi96,SaAnSh99,NgRoSh00}.  
In the absence of multivalent
counterions, DNA inside solution  has extended configuration.
When the concentration of multivalent salt is
slowly raised, 
there comes a point at which
the gain in electrostatic energy due to polyion-$\alpha$-ion
association  overcomes the 
entropic loss 
due to $\alpha$-ion confinement in the vicinity of the polyions.  
The condensed monovalent counterions are then released into solution
and are replaced by the
polyvalent $\alpha$-ions.  After the critical number of
$\alpha$-ions is associated to the polyion, the interaction
between the separate segments of the DNA  becomes
attractive.  This plus the intrinsic rigidity of the molecule
drives its condensation into toroidal bundles.  As the
concentration of multivalent ions in solution increases further,
the DNA-$\alpha$-ion complex becomes overcharged.  When
the effective charge of the polyion reaches 
$Z_{eff}^u$, the interaction between the
segments of the DNA becomes, once again, repulsive 
and the bundles re-dissociate.
Clearly this simple model is incapable of accounting for all
the intricacies of the DNA condensation, nevertheless
it sheds a lot of light on the 
role of  electrostatic correlations 
in this interesting and important phenomenon.

The calculations above were presented for a very idealized model
of interacting lines of charge.  It is quite simple
to modify the theory to account for finite polyion 
diameter.  This modification, however, does not significantly affect 
the predictions of the theory.  Attraction appears at small separations
between the polyion surfaces --- about $7$ \AA --- after the  
critical number of $\alpha$-ions is 
condensed onto the polyions~\cite{DiCaLe01}.  
We find that for macromolecules of finite diameter less counterions
are needed to induce attraction than for the two lines of
equivalent charge density~\cite{DiCaLe01}. Furthermore, the 
charge-charge correlations along the polyion
are of very short range~\cite{DiCaLe01,DeArHo02}, 
showing absence of any long-range
order between the condensed counterions, contrary to 
the earlier speculations~\cite{RoBl96,GrMaBr97,Sh99}.

\begin{figure}  
\begin{center}
\includegraphics[width=10cm,angle=270]{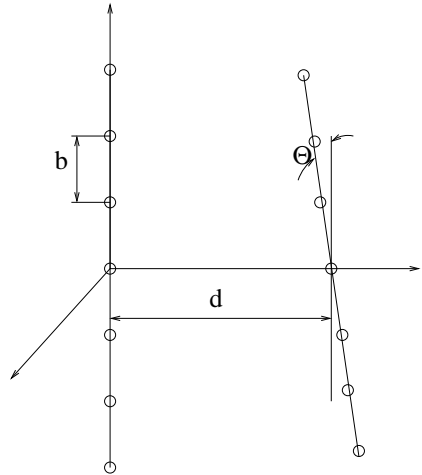}
\end{center}
\caption{Two rodlike polyions of $Z=7$.}
\label{Fig9}
\end{figure}

\begin{figure}  
\begin{center}
\psfrag{f}{${\cal F}$}
\includegraphics[width=8cm,angle=270]{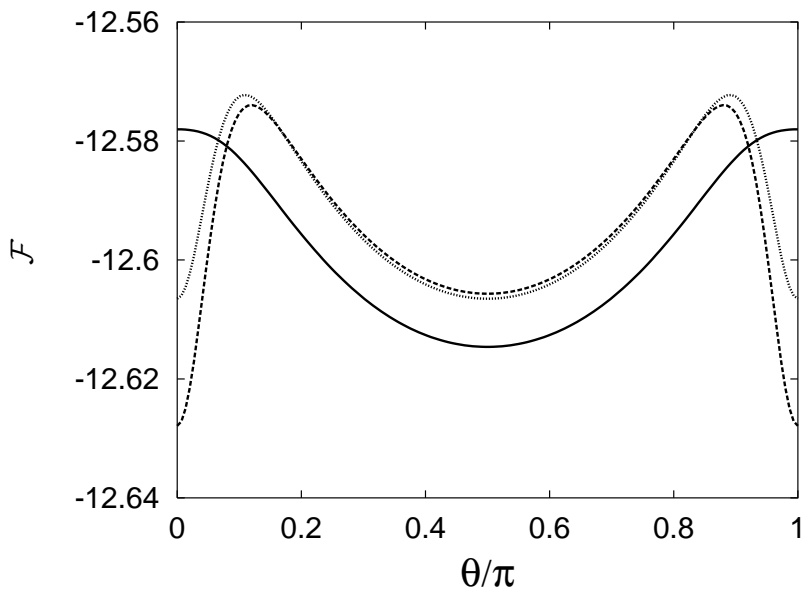}
\end{center}
\caption{Reduced free energy ${\cal F}$ as a function 
of the angle $\theta$ for
$Z=9$, $n=4$, $\alpha=2$, $\xi=2$. 
The curves shown are for: $x=2$ (full
line), $x=1.3676$ (dotted line), and $x=1.3$ (dashed line).
Notice that the free energies for $\theta=0$ and $\theta=\pi/2$ are
equal in the second case~\cite{StLeAr02}.}
\label{Fig10}
\end{figure}

A very interesting experimental feature of rigid polyelectrolyte
solutions
containing multivalent counterions is that the 
correlation-induced attraction does not lead to 
phase separation~\cite{HaLi00}.  Instead 
rodlike polyions associate into bundles of well defined thickness.
The precise  bundle morphology depends on the persistence
length of  the polyions, 
all the bundles, however, tend to have a well 
defined cross-sectional diameter. What can account for this
curious phenomenon?  Correlation induced attraction
should favor an unconstrained growth of bundles~\cite{Sh99}.  In this
respect the situation is very similar to that of an 
ionic crystal whose electrostatic energy 
is negative and is unbounded from bellow.  
What then cuts off the bundle size?  
The question is still not fully settled, but
the indications are that the size of a bundle is controlled by the 
kinetics of its formation~\cite{HaLi99}.  To understand this better, 
lets, once again, consider 
our simple model of two lines of charge
with $n_l<n<Z/\alpha$ condensed $\alpha$-ions~\cite{StLeAr02}.
The polyions are located on
two parallel planes separated by a distance $d$.
They are free to rotate in their respective planes. 
The relative angle between the two lines is 
$\theta$, with  $\theta=0$ corresponding to two  
parallel polyions, Fig.~\ref{Fig9}.   
It is convenient to
define  the adimensional free energy as
\begin{equation}
{\cal F}= -\frac{1}{\xi}\ln Q \;,
\label{179}
\end{equation}
\newpage
\noindent where  $Q$ is the partition function for the 
two polyions at fixed separation  and
angular orientation and $\xi$ is the Manning parameter.
In Fig. \ref{Fig10} the reduced free energy is plotted as
a function of $\theta$ for various separations between
the polyions.
The free energy has two extrema at $\theta=0$ and  $\theta=\pi/2$, 
corresponding to 
the parallel and the perpendicular orientation between the polyions. 
In general, in situations when
the attraction appears, 
the perpendicular configuration has the lowest free
energy at large separations between the polyions,  
while at small distances the parallel
configuration minimizes the free energy of the model.  However, 
in some cases, a reentrant behavior is observed.  The
perpendicular configuration is the global minimum for both large and small
distances, while the parallel configuration minimizes the free energy
at intermediate distances.

At large separations the like-charged polyion-counterion
complexes repel and are perpendicular to one another.  As the distance
between the polyions is decreases, the global minimum passes 
from $\theta=\pi/2$
to $\theta=0$, with  $\theta=\pi/2$ becoming metastable.  
In order for polyions to form a bundle they
must align, which means they  
have to overcome an activation barrier which separates the
metastable minimum at $\theta=\pi/2$ from the global
minimum at $\theta=0$. It has been suggested
that the height of the barrier  grows with the number of 
polyions which are already in the bundle~\cite{HaLi99}.  There 
comes a point when the barrier  becomes 
so high that the thermal fluctuations are unable to overcome it
and no new polyion can join the bundle.  This puts an 
end to the bundle growth.

\subsection{Counterion polarization}

\begin{figure}  
\begin{center}
\psfrag{f}{${\cal F}$}
\includegraphics[width=8cm]{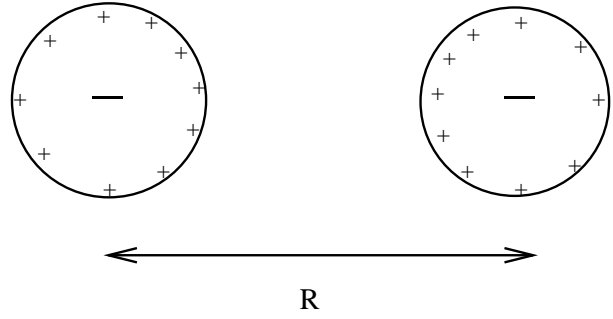}
\end{center}
\caption{Two polyions with condensed counterions.  Note that
the net charge $Z_{eff}$ of one complex polarizes the counterions
of the other complex.}
\label{Fig11}
\end{figure}

The correlation-induced attraction between the 
polyions discussed in the previous
section is short ranged, decaying exponentially
with the separation between the polyions.  The line
of charge model is, however, too simple to account
for the counterion polarization~\cite{Be00}.  Consider for example
a rigid cylinder-like polyion or a spherical colloidal
particle with a layer of condensed counterions.  Suppose
that there is no overcharging so that only some fraction
of the  polyion charge is neutralized by the condensed counterions.
The electric field  produced by one 
complex  polarizes the counterions of 
the other  complex, Fig. \ref{Fig11}.  
This mechanism certainly
provides an attractive component to the  interaction
potential~\cite{Le99a}.  However, can this
induced attraction overcome the monopolar repulsion
from the uncompensated colloidal charge $Z_{eff}$?  To understand this
lets consider two colloidal particles or globular
proteins separated by a distance $R$ inside an electrolyte solution
with  Debye length $\xi_D$. Each  polyion has
radius $a$, charge $Z$, and $n$ condensed $\alpha$-ions.
The leading order interaction between
the two macromolecules at separation $R>\xi_D$ is of $DLVO$ form
\begin{equation}
\label{180}
V_{DLVO}(R)=Z_{\rm eff}^2 q^2 \theta^2(\kappa a) \frac{e^{-\kappa R}}{\epsilon R},
\end{equation}
where $Z_{eff}=Z-\alpha n$
The major corrections come from two effects.
First, presence of polyions produces holes in the ionic atmosphere.  
The charge of a hole is
minus the charge of the counterions excluded 
from the region of space occupied by the 
macromolecule~\cite{LiLeFi94,FiLeLi94}. For example, the charge of the
hole produced by the polyion ${\cal P}_2$ in the ionic
atmosphere polarized by the polyion ${\cal P}_1$ is
\begin{equation}
\label{181}
Q_{h,2}(R) \approx -\frac{4 \pi a^3 \rho_q(R)}{3} \;,
\end{equation}  
where the charge density at position $R$ is given by the
familiar Debye-H\"uckel expression,
\begin{equation}
\label{182}
\rho_q(R)=-\frac{\epsilon \kappa^2}{4 \pi} \phi(R)=
 Z_{\rm eff}q \kappa^2\theta(\kappa a) \: \frac{e^{-\kappa R}}{4 \pi R} \;.
\end{equation}  
Use of the linearized Debye-H\"uckel theory is justified
by the fact that in Eq.~(\ref{182}) enters 
the renormalized effective charge.
The electrostatic energy of the polyion-hole interaction
is obtained using the charging 
process  in which $q \rightarrow \lambda q$  and
\begin{equation}
\label{183}
V_h(R)= Q_{h,2}(R) \int_0^1 d\lambda \phi(R;\lambda q)=\frac{1}{6} \kappa^2 a^3
 Z_{\rm eff}^2 q^2 \theta^2(\kappa a)\: \frac{e^{-2 \kappa R}}{\epsilon R^2}\;,
\end{equation}  
at fixed  $\kappa$.
The second major correction to the DLVO potential is the result of 
polarization of the condensed layer of counterions surrounding the polyion.
Thus, the electric field produced by  
the complex ${\cal P}_1$ induces  a dipole moment
in the complex  ${\cal P}_2$, and vice-versa.   The
local surface charge density of  ${\cal P}_2$ is
\begin{equation}
\label{184}
\sigma({\bf r})= -\sigma_-+ \sigma_+e^{-\beta \alpha q \phi({\bf r})} \;, 
\end{equation}
where 
\begin{equation}
\label{185}
\sigma_-=\frac{Zq}{4 \pi a^2} \;, 
\end{equation}
\begin{equation}
\label{186}
\sigma_+=\frac{\alpha n q}{4 \pi a^2} \;, 
\end{equation}
and $\phi({\bf r})$ is the local electrostatic potential felt by the
condensed counterions at position ${\bf r}$.  In the weak coupling limit
$\Gamma < 1$ the exponential in Eq.~(\ref{184}) can be linearized 
yielding the surface charge distribution
\begin{equation}
\label{187}
 \sigma({\bf r})=\Delta \sigma - 
\frac{\epsilon \phi({\bf r})}{2 \pi \lambda_{GC}}\;, 
\end{equation}
where $\Delta \sigma=\sigma_+-\sigma_-$ and the
Gouy-Chapman length is
\begin{equation}
\label{188}
\lambda_{GC}=\frac{\epsilon}{2 \pi \sigma_+ \beta \alpha q} \;. 
\end{equation}
It is a straightforward
calculation in electrostatics to show that 
the induced dipole moment of
the complex ${\cal P}_2 $ is
\begin{equation}
\label{189}
{\bf p}=\frac{2 a^4 \epsilon {\bf E}}{(3\lambda_{GC}+2 a)} \;.
\end{equation}
The  electric field 
produced by ${\cal P}_1$
at distance $R$ from its center is
\begin{equation}
\label{190}
{\bf E}({\bf R})=Z_{\rm eff} q \theta(\kappa a) \:\frac{e^{-\kappa R} 
(1+\kappa R) {\bf R}}{\epsilon R^3} \; .
\end{equation}
The electrostatic energy of the induced dipole of ${\cal P}_2$ inside the
electric field produced by ${\cal P}_1$ is obtained using
the charging process
\begin{equation}
\label{191}
V_p=\int_0^1 d \lambda {\bf p}(\lambda q) \cdot {\bf E}(\lambda q) \;,
\end{equation}
where it is important to remember that 
$\lambda_{GC}(\lambda q)=\lambda^{-2} \lambda_{GC}(q)$. 
The same argument can be equally well applied to the
interaction of ${\cal P}_2$ with the hole
produced by ${\cal P}_1$ and to polarization of ${\cal P}_1$
by the electric field produced by  ${\cal P}_2$.
Summing up all of these contributions leads to
\begin{equation}
\label{192}
W(R)=Z_{\rm eff}^2 q^2\theta^2(\kappa a) \frac{e^{-\kappa R}}{\epsilon R}- 
Z_{\rm eff}^2 q^2 \kappa^2 a^3 \theta^2(\kappa a)\: 
h\left(\frac{a}{\lambda_{GC}}\right) \:
\frac{e^{-2 \kappa R}}{\epsilon R^2} ,
\end{equation}
where the scaling function $h(x)$ is, 
\begin{equation}
\label{193} 
h(x)=\frac{2}{3}-\frac{3}{2 x} \:\ln\left(1+\frac{2 x}{3}\right). 
\end{equation}
The correction to the $DLVO$ potential is repulsive 
(hole dominated) for $ a/\lambda_{GC} < 1.716...$ 
and is attractive (dipole dominated) for $a/\lambda_{GC} > 1.716...$. 
We note, however, that it is always 
``doubly'' screened and is, therefore,
smaller than the leading order $DLVO$ term
for separations $R>\xi_D$~\cite{NiPa71}.  
At shorter distances the approximations
used to arrive at Eq.~(\ref{192}) fail, and
the correlations between the
condensed counterions must be explicitly taken into account~\cite{Le99a}.
For multivalent counterions, the electrostatic 
correlations result in  short-ranged attraction, 
similar to the one discussed in the previous section.

\section{Conclusions}

We have explored the role of electrostatic correlations in systems
ranging from classical plasmas to molecular biology. We saw how
the positional correlations between the ions of an electrolyte can 
result in a thermodynamic instability.  We also saw how the 
strong correlations between the polyions and the counterions
lead to colloidal charge renormalization 
which stabilizes deionized
suspensions against a phase separation.  For two-dimensional 
plasmas electrostatic correlations are responsible for the 
metal-insulator transition.
The critical behavior of 
superfluid $^4$He films,
the roughening transition of crystal interfaces~\cite{ChWe76}, 
the melting of 
two-dimensional solids, and the criticality
of XY magnets~\cite{Ne83} 
are all governed by the ``electrostatic'' interactions between the
topological defects (charges). 
Nature has learned to 
take the full advantage of
electrostatic correlations to efficiently
package huge amounts of genetic material into tiny regions 
of space.  

Throughout this review we have come to rely on some simple
models in order to understand complex physical phenomena.  
While these models are often sufficient to grasp 
the underlying physics,
it is quite easy to push the models too far.  
This is particularly the
case when one deals with specific structural 
properties of biomolecules~\cite{GiRaFi85,PeWaJo98,SpBa98}.  
As soon as the  length scales on the order of
few angstroms become important, approximation of water as a uniform
dielectric medium is no longer sufficient~\cite{AuWe98,AuLoWe96}.  
Under these conditions reliance on simple models, which treat
macromolecules and solvent as 
dielectrics, is probably 
no more than a wishful thinking.  
A careful path must be threaded between 
the simplification and the over-simplification.

\section{Acknowledgements}
 
I am grateful to J.J. Arenzon, M.C. Barbosa, A. Diehl, 
M.E. Fisher, J.E. Flores-Mena, 
P. Kuhn, X.-J. Li, J. Stilck, and M.N. Tamashiro 
who have collaborated closely with me on  
topics covered in this review.  
Special thanks are due to Michael Fisher and 
Shubho Banerjee who have kindly
provided  me with Fig. \ref{Fig0} which compares the
predictions of 
various theories of asymmetric electrolytes.  
Last but not least, I would like to 
thank Eric Westhof without whose
encouragement this review would not have been written.

\newpage

\end{document}